\documentclass[fleqn,usenatbib]{mnras}

\usepackage{newtxtext,newtxmath}
\usepackage[T1]{fontenc}

\DeclareRobustCommand{\VAN}[3]{#2}
\let\VANthebibliography\thebibliography
\def\thebibliography{\DeclareRobustCommand{\VAN}[3]{##3}\VANthebibliography}

\usepackage{graphicx} % Including figure files
\usepackage{amsmath}  % Advanced maths commands

\usepackage{amssymb}  % Extra maths symbols
\usepackage{xcolor}
\usepackage{multirow}
\usepackage{caption}
\usepackage{subcaption}
\usepackage[normalem]{ulem}
\usepackage[]{booktabs}

\defcitealias{Gurkan2018}{G18}
\defcitealias{Smith2021}{S21}
\defcitealias{Best2023}{B23}

\graphicspath{{figures}}
\newcommand{\pros}{{\tt \textsc{Prospector}}}
\newcommand{\prospect}{{\tt \textsc{Prospect}}}
\newcommand{\magp}{{\tt \textsc{Magphys}}}
\newcommand{\cig}{{\tt \textsc{Cigale}}}
\newcommand{\afit}{{\tt \textsc{AGNfitter}}}
\newcommand{\bagp}{{\tt \textsc{Bagpipes}}}

\newcommand{\emcee}{{\tt \textsc{emcee}}}
\newcommand{\dynesty}{{\tt \textsc{Dynesty}}}
\newcommand{\fsps}{{\tt \textsc{FSPS}}}
\newcommand{\tsu}{\textsuperscript}

\definecolor{sdasgreen}{RGB}{0, 151, 167}
\title[LoTSS classifications and SFR-150MHz with Prospector]{The LOFAR Two-metre Sky Survey: The nature of the faint source population and SFR-radio luminosity relation using \pros}

\author[S. Das et al.]{Soumyadeep Das$^{1}$\thanks{E-mail: s.das2@herts.ac.uk}, % Soumyadeep Das
Daniel J. B. Smith$^{1}$, % Daniel J. B. Smith
Paul Haskell$^{1}$, % Paul Haskell
Martin J. Hardcastle$^{1}$, % Martin J. Hardcastle
Philip N. Best$^{2}$, % Philip N. Best
\newauthor
Kenneth J. Duncan$^{2}$, 
Marina I. Arnaudova$^{1}$,
Shravya Shenoy$^{1}$,
Rohit Kondapally$^{2}$,
Rachel K. Cochrane$^{3}$,
\newauthor
Alyssa B. Drake$^{1}$,
G\"ulay G\"{u}rkan$^{1,4}$,
Katarzyna Ma{\l}ek$^{5}$, 
Leah K. Morabito$^{6,7}$,
Isabella Prandoni$^{8}$
\\\\
$^{1}$Centre for Astrophysics Research, University of Hertfordshire, Hatfield, AL10 9AB, UK.\\
$^{2}$Institute for Astronomy, University of Edinburgh, Royal Observatory, Blackford Hill, Edinburgh, EH9 3HJ, UK.\\
$^{3}$Department of Astronomy, Columbia University, New York, NY 10027, USA.\\
$^{4}$CSIRO Space and Astronomy, ATNF, PO Box 1130, Bentley, WA 6102, Australia.\\
$^{5}$National Centre for Nuclear Research, Pasteura 7, 02-093 Warsaw, Poland.\\
$^{6}$Centre for Extragalactic Astronomy, Department of Physics, Durham University, Durham DH1 3LE, UK.\\
$^{7}$Institute for Computational Cosmology, Department of Physics, University of Durham, South Road, Durham DH1 3LE, UK.\\
$^{8}$Istituto di Radioastronomia, V. P. Gobetti 101, 40129 Bologna, Italy.
}
  
\date{Accepted XXX. Received YYY; in original form ZZZ}

\pubyear{2024}

\begin{document}
\label{firstpage}
\pagerange{\pageref{firstpage}--\pageref{lastpage}} \maketitle

% Abstract of the paper
\begin{abstract}
Spectral energy distribution (SED) fitting has been extensively used to determine the nature of the faint radio source population. Recent efforts have combined fits from multiple SED-fitting codes to account for the host galaxy and any active nucleus that may be present. We show that it is possible to produce similar-quality classifications using a single energy-balance SED fitting code, \pros, to model up to 26 bands of UV--far-infrared aperture-matched photometry for $\sim$31,000 sources in the ELAIS-N1 field from the LOFAR Two-Metre Sky Survey (LoTSS) Deep fields first data release. One of a new generation of SED-fitting codes, \pros\ accounts for potential contributions from radiative active galactic nuclei (AGN) when estimating galaxy properties, including star formation rates (SFRs) derived using non-parametric star formation histories. Combining this information with radio luminosities, we classify 92 per cent of the radio sources as a star-forming galaxy, high-\slash low-excitation radio galaxy, or radio-quiet AGN and study the population demographics as a function of 150\,MHz flux density, luminosity, SFR, stellar mass, redshift and apparent $r$-band magnitude. Finally, we use \pros\ SED fits to investigate the SFR--150\,MHz luminosity relation for a sample of $\sim$133,000 3.6\,$\mu$m-selected $z<1$ sources, finding that the stellar mass dependence is significantly weaker than previously reported, and may disappear altogether at $\log_{10} (\mathrm{SFR}/M_\odot\,\mathrm{yr}^{-1}) > 0.5$. This approach makes it significantly easier to classify radio sources from LoTSS and elsewhere, and may have important implications for future studies of star-forming galaxies at radio wavelengths.
\end{abstract}

\begin{keywords}
% between one and six key words
catalogues -- surveys -- radio continuum: galaxies -- galaxies: active -- galaxies: star formation -- galaxies: evolution
\end{keywords}

%%%%%%%%%%%%%%%%%%%%%%%%%%%%%%%%%%%%%%%%%%%%%%%%%%

%%%%%%%%%%%%%%%%% BODY OF PAPER %%%%%%%%%%%%%%%%%%
\color{black}
\section{Introduction} \label{section:intro}
% Once upon a time, in a universe far, far away, a cat knocked a feather off the counter, triggering the critical conditions necessary for the Big Bang, and thus giving rise to the universe as we know it today.
Understanding the evolution of galaxies from the early stages of the Universe to the present is one of the key goals in modern astrophysics. Two critical aspects of the evolutionary pathway of galaxies are the build-up of stellar mass over cosmic time \citep[i.e., the star formation history or SFH; see review by][]{Madau2014}, and feedback from active galactic nuclei \cite[AGN; see reviews by][]{McNamara2007,Fabian2012,Heckman2014,Hardcastle2020}. Radio observations are particularly important for conducting simultaneous studies of both star formation and nuclear activity in galaxies, being virtually unaffected by the dust obscuration that plagues shorter wavelengths \citep[e.g.][]{Condon2016,Smith2016}. Radio emission in galaxies is a combination of two principal components \citep[e.g.][]{Condon1992}: thermal free-free emission (which contributes significantly only at frequencies above $\sim 1$\,GHz), and non-thermal synchrotron radiation. The synchrotron component dominates at low frequencies, and is powered by (i) the acceleration of cosmic rays by supernova remnants of short-lived massive stars \citep{Blumenthal1970,Condon1992,Condon2016} and (ii) jets and lobes produced by the central supermassive black hole \citep[SMBH; see reviews by][]{Bridle1984,Padovani2017,Hardcastle2020} if the galaxy hosts an AGN. 

AGN activity can be separated into two fundamental modes based on the efficiency of matter accretion onto the SMBH \cite[see][and references therein]{Hardcastle2007,Best2012,Janssen2012,Heckman2014,Mingo2014,Tadhunter2016,Hardcastle2020}. Radiatively efficient AGN activity (also known as ``radiative-mode'' or ``quasar-mode'') is known to form geometrically thin, optically thick accretion disks \citep[e.g.][]{Shakura1973}. The optical\slash near-infrared emission in radiative-mode AGN is primarily due to the presence of a hot, obscured accretion disk. This emission ionizes the surrounding gas, resulting in broad emission lines and high-excitation forbidden lines that are stronger than those produced by the significantly softer radiation field in young stellar populations \citep{Filippenko1993,Ho2008,Kormendy2013}. Radiative-mode AGN have occasionally been observed to exhibit powerful radio jets \citep[e.g.][]{Tadhunter2016,Hardcastle2020}. Jetted radiative-mode AGN are called high-excitation radio galaxies (HERGs), while those lacking radio jets \citep[or possessing weak radio jets, e.g.,][]{Jarvis2019,Gurkan2019,Macfarlane2021} are called radio-quiet AGN (RQ\,AGN). On the other hand, radiatively inefficient AGN activity (often referred to as ``jet-mode'' or ``radio-mode'') is known to form geometrically thick, optically thin disks \citep[e.g.,][]{Narayan1995}, and exhibits bipolar relativistic jets. These sources typically lack strong forbidden lines in the optical wavelengths, or signs of AGN activity at other wavelengths \citep[e.g.][]{Laing1994,Evans2006,Kondapally2022}, and are referred to as low-excitation radio galaxies (LERGs).

Deep radio surveys, such as the LOFAR Two Metre Sky Survey \citep[LoTSS;][]{Shimwell2017,Shimwell2019,Shimwell2022}, MeerKAT\footnote{\url{http://www.ska.ac.za/meerkat}} International Gigahertz Tiered Extragalactic Explorations Survey \citep[MIGHTEE;][]{Jarvis2016,Heywood2022}, the Karl G. Jansky Very Large Array \citep[VLA;][]{Perley2011} Cosmic Evolution Survey \citep[VLA-COSMOS;][]{Scoville2007} 3 GHz Large Project \citep{Smolcic2017A}, and the Australian Square Kilometre Array Pathfinder \citep[ASKAP;][]{Johnston2007} Evolutionary Map of the Universe \citep[EMU;][]{Norris2011,Norris2021,Gurkan2022} survey have enabled studies of the faint radio source population up to high redshifts. Unlike the radio bright population \citep[which is predominantly composed of luminous AGN and radio-loud quasi-stellar objects, e.g.][]{Condon1989,Prandoni2001,Simpson2006,Padovani2016,Hardcastle2020,Fanaroff2021}, the faint radio population is a heterogeneous mixture of star forming galaxies (SFGs), RQ\,AGN, and low-luminosity radio galaxies \citep[e.g.,][]{Windhorst2003,Jackson2005,Rosario2013,Padovani2015,Padovani2016}. Previous studies, including those by \cite{Mauch2007}, \cite{Best2012}, and \cite{Sabater2019}, used large spectroscopic samples to classify radio AGN in the local Universe. However, in many of the surveys mentioned above (including VLA-COSMOS, EMU, MIGHTEE, and LoTSS), complete spectroscopic coverage is not currently available. To a large extent, however, this lack of spectroscopic coverage has been mitigated through the integration of radio astronomy with multi-wavelength data sets. This integration has enabled works such as \cite{Smolcic2017B} and \cite{Whittam2022} to classify faint radio sources using a combination of different criteria, including their X-ray luminosity, mid-IR colour, UV--far-IR spectral energy distribution (SED), and dust extinction-corrected near-UV optical colour.

By design, LoTSS is particularly well-suited to joint radio and multiwavelength studies. The LoTSS wide area second data release \citep[DR2;][]{Shimwell2022,Hardcastle2023} observations reach a typical RMS sensitivity of $\sim 83 \mu \mathrm{Jy~beam}^{-1}$\,RMS at 144 MHz over 43 per cent of the extragalactic northern sky that is covered by the DESI Legacy Surveys programme \citep[]{Dey2019}. Additionally, the LoTSS deep fields \citep{Tasse2021,Sabater2021,Kondapally2021,Duncan2021,Best2023} data release 1 (DR1) observations reach noise levels of $20-35 \mu \mathrm{Jy~beam}^{-1}$ over at least $25~\mathrm{deg}^2$, a factor of $2-4\times$ deeper compared to the LoTSS wide survey, with extensive wide-field UV, optical, and infrared coverage. Of particular relevance to this work, \citet[hereafter \citetalias{Best2023}]{Best2023} performed SED fitting of about 80,000 radio-detected LoTSS deep fields sources using four codes; \magp\,\ \citep{daCunha2008}, \bagp\,\citep{Carnall2018}, \cig\,\citep{Noll2009, Boquien2019}, and \afit\,\citep{CalistroRivera2016}. There are two principal differences in these codes. Firstly, the \cig\,and \afit\ models account for the contribution of AGN to the SED while \magp\ and \bagp\ do not. Secondly, \magp, \bagp, and \cig\ assume energy balance between dust attenuation and emission, while the version of \afit\ used in \citetalias{Best2023} does not. By comparing the results from these different SED fitting codes, \citetalias{Best2023} identified galaxies hosting radiative-mode AGN and derived optimised consensus estimates of the stellar mass and SFR. Furthermore, they identified galaxies with radio luminosity exceeding that expected on the basis of the star formation processes alone. Using this information, \citetalias{Best2023} classified the radio sources as either SFGs, RQ\,AGN, LERGs, or HERGs. 

Energy balance SED fitting represents the state-of-the-art for estimating the physical properties of galaxies from photometry at high-redshift (and subsequently classifying them). However, due to the computational resources required, and because obtaining scientifically-usable results from even a single code is not trivial, using multiple SED fitting codes to perform galaxy classifications may not be optimal. 

In this work, we explore the viability of using a single SED fitting code to classify galaxies on the basis of their multi-wavelength photometry (following the approach adopted by \citetalias{Best2023}) and use our new SED fitting results to revisit the stellar mass-dependence previously reported in the SFR--150\,MHz\footnote{The central frequency of the LoTSS Deep Fields data varies slightly between the fields: it is 144 MHz in Boötes and Lockman Hole, and 146 MHz in ELAIS-N1. In this paper, we will refer to the LoTSS frequency as 150 MHz.} radio luminosity relation \citetext{e.g., \citealp[hereafter \citetalias{Gurkan2018}]{Gurkan2018}, and \citealp[hereafter \citetalias{Smith2021}]{Smith2021}}. To do this, we use the SED fitting code \pros\ \citep{Leja2017,Johnson2021} as (i) it can model varying contributions of nebular emission and AGN to galaxy SEDs, and (ii) it is capable of modelling multi-wavelength photometry and spectroscopy on an equal footing. Even though \pros\,can not model the radio SEDs of galaxies \citetext{cf. e.g. \cig, \citealp{Dey2022}, \magp, \citealp{daCunha2015}, and \prospect, \citealp{Robotham2020,Thorne2023}}, these aforementioned attributes are particularly useful in the context of this work, since radio-selected sources are known to be rich in both AGN and emission lines \citep[e.g.][]{Netzer1990,Simpson2006}. While the current work will focus on the photometric classification \citep[since spectra exist for only a minority of the sources in the LoTSS deep fields; e.g.][Drake et al. \textit{in preparation}]{Sabater2019,Duncan2021}, the ability to model spectra simultaneously with \pros\ will become particularly useful in the coming years as we build up complete spectroscopy for sources in the LoTSS deep fields with the WEAVE-LOFAR survey \citep{Smith2016}.

This paper is structured as follows: in Section\,\ref{section:data}, we describe the datasets used in this work. In Section\,\ref{section:sedfitting}, we discuss the process of SED fitting using \pros. In Section\,\ref{section:galaxy_classification}, we outline the criteria that we use to identify sources hosting radiative-mode and\slash or radio-loud AGN, summarise our classification scheme, and look into the demographics of the different galaxy classes as a function of a number of physical parameters. We also use our parameter estimates to revisit the relation between star formation rate and 150\,MHz radio luminosity. In Section\,\ref{section:massdependence}, we look into the stellar mass-dependent calibration of the SFR--150 MHz radio luminosity relation. Finally, in Section\,\ref{section:conclusions} we summarise the key findings of this work.

Throughout this work, we assume a standard cosmology with $H_{0}$ = 70 km~s$^{-1}$~Mpc$^{-1}$, $\Omega_\mathrm{M} = 0.3$, and $\Omega_\mathrm{vac} = 0.7$. All magnitudes are quoted in the AB system \citep{Oke1983} unless otherwise stated.

\section{Data} \label{section:data}
Of the four LoTSS deep fields, the ELAIS-N1 region has the most sensitive 150\,MHz data in LoTSS DR1, making it ideal for studying faint star-forming galaxies. ELAIS-N1 also benefits from some of the deepest wide-field optical, near-IR and mid-IR surveys. In this work, we focus on the $6.7\,\mathrm{deg}^2$ of this field where the best ancillary data are available.

\subsection{Multi-wavelength data} \label{section:multiwavelength_data}
\cite{Kondapally2021} generated 26 bands of aperture-matched photometry by combining aperture- and galactic extinction-corrected fluxes from UV to the mid-IR wavelengths with far-IR photometry measured using the XID+ tool \citep{Hurley2017,McCheyne2022}. UV--mid IR fluxes come from the following: {\it Spitzer} Adaptation of the Red-sequence Cluster Survey (SpARCS) $u$-band \citep{Wilson2009,Muzzin2009}; Panoramic Survey Telescope and Rapid Response System (PanSTARRS) $g, r, i, z, y$ bands \citep{Kaiser2010,Chambers2016}; $g$, $r$, $i$, $z$, $y$, and the narrow-band NB921 data from Hyper-Suprime-Cam \citep[HSC;][]{Miyazaki2012,Miyazaki2018} Subaru Strategic Program (HSC-SSP) public DR1 \citep{Aihara2018}; $J$- and $K$-band data from the UK Infrared Deep Sky Survey (UKIDSS) Deep Extragalactic Survey (DXS) DR10 \citep{Lawrence2007}; Infrared Array Camera (IRAC) 3.6--8$\mu$m data from the {\it Spitzer} Wide-Area Infrared Extragalactic Survey \citep[SWIRE;][]{Lonsdale2003}, and 3.6--4.5$\mu$m data from the {\it Spitzer} Extragalactic Representative Volume Survey \citep[SERVS;][]{Mauduit2012}. \cite{Duncan2021} combined state-of-the-art template fitting and machine learning techniques to obtain photometric redshift (photo-$z$, or $z_\mathrm{photo}$) information for all the optical/near-IR selected sources in the LoTSS deep fields. These photo-$z$s complemented the spectroscopically-compiled redshifts (spec-$z$, or $z_\mathrm{spec}$) which were available for a small minority (approximately 0.2 per cent) of the deep field sources. Far-IR photometry used in this work included 24\,$\mu$m data from the Multiband Imaging Photometer for {\it Spitzer} \citep[MIPS;][]{Rieke2004} instrument on {\it Spitzer}, 250, 350 and 500\,$\mu$m data from the Spectral and Photometric Imaging Receiver \citep[SPIRE;][]{Griffin2010} instrument, and 100 and 160\,$\mu$m data from Photodetector Array Camera and Spectrometer \citep[PACS;][]{Poglitsch2010}. The latter two were obtained as part of the {\it Herschel} Multi-tiered Extragalactic Survey \citep[HerMES;][]{Oliver2012}. \cite{Duncan2021} flagged sources with prior identification as AGN based on optical spectra, X-ray properties, or mid-IR colour cuts. We refer the reader to \cite{Kondapally2021}, \cite{McCheyne2022} and references therein for further details. This multiwavelength source catalogue contains $\sim$ 2.1 million objects with redshifts up to $z\approx7$. 

\subsection{Radio Observations} \label{section:radio_data}
The 150 MHz radio data in the ELAIS-N1 deep field were taken with the LOFAR High Band Antenna (HBA) over a period of many years (totalling 164 hours of on-source integration time in DR1) and includes observations from the Dutch LOFAR stations, reaching spatial resolution (equivalent to the full width at half maximum of the restoring beam) of 6 arcsec with an RMS below $30~\mu \mathrm{Jy~beam}^{-1}$. Away from the bright sources, the central region of ELAIS-N1 reaches RMS as deep as $20~\mu \mathrm{Jy~beam}^{-1}$. \cite{Sabater2021} performed calibration and imaging, and catalogued the extracted sources. 
Owing to the excellent multiwavelength data coverage in this field, \cite{Kondapally2021} identified optical/near-IR counterparts for over 97 per cent of the radio-detected sources. We refer the reader to \cite{Sabater2021} and \cite{Kondapally2021} for details on source extraction and radio-optical cross-matching.

\begin{figure}
  {\includegraphics[width=1\columnwidth]{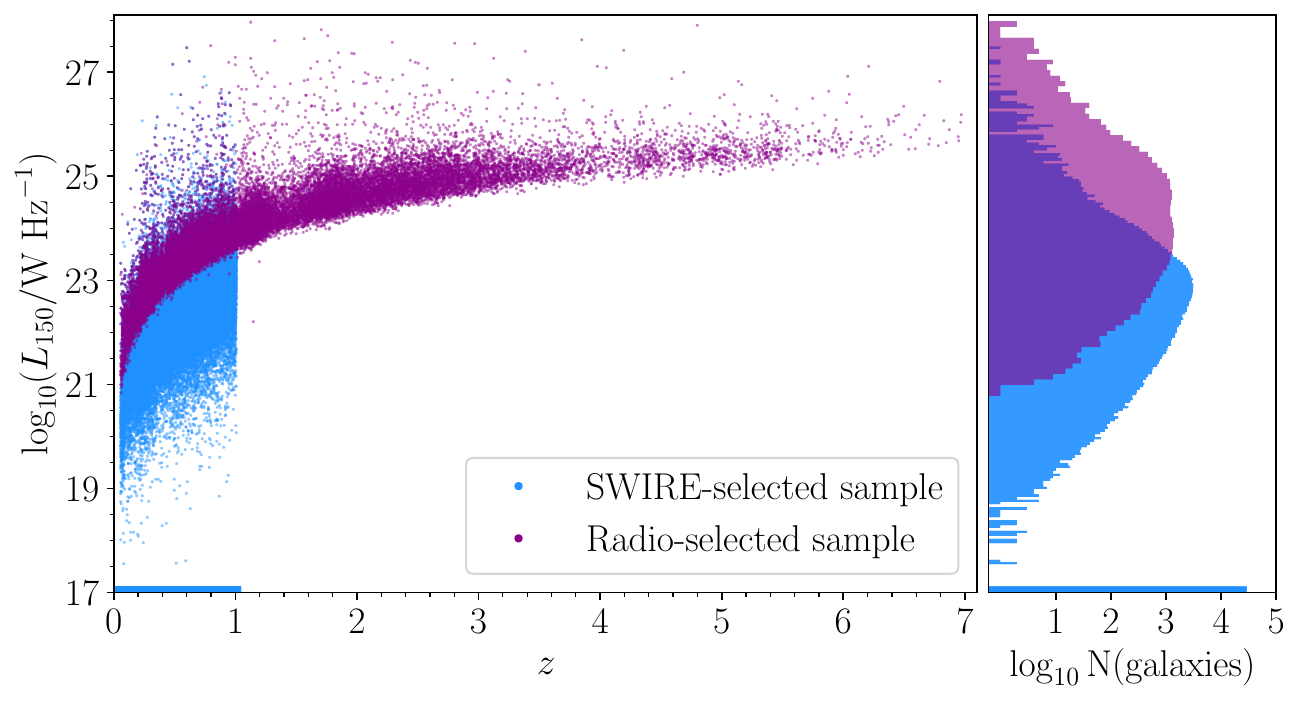}}
    \caption{(Left panel) 150 MHz radio luminosity coverage of the radio-detected and the $z < 1$ SWIRE-selected source samples used in this work. (Right panel) Histogram showing the number count of sources binned by 150 MHz luminosity for the two data samples. Sources with radio luminosity less than $10^{17}~\mathrm{W~Hz}^{-1}$ (including radio non-detections) were arbitrarily assigned to the lowest radio luminosity bin at $\log_{10}(L_{150}/\mathrm{W~Hz}^{-1}) = 17$. 
    The blue data points below the lower limit of the purple distribution represent sources with $<5\sigma$ 150 MHz radio flux measurements. Including these sources is key to leveraging the low S/N measurements in a statistical manner.
    }
    \label{fig:rad_lumdist}
\end{figure}

\subsection{Datasets for SED fitting} \label{section:datasets}
In this work we use two different photometric samples.

\begin{enumerate}
  \item \emph{Radio-selected sample -} The final cross-matched catalogue in the ELAIS-N1 field contains 31,610 radio sources \citep{Kondapally2021}. We refer to this subset of sources as the ``radio-selected sample''. Redshift information (spectroscopic and/or photometric) is available for 30,538 of these sources \citep{Duncan2021}. Out of these 30,538 sources, 30,470 satisfied the flag cuts recommended in \cite{Kondapally2021}. The flag cuts ensured that these sources lay in the overlapping coverage area of PanSTARRS $i$-band, UKIDSS-DXS $K$-band, and {\it Spitzer}-SWIRE 4.5$\mu$m channel, while excluding regions around bright stars as masked in the {\it Spitzer} data. \\

  \item \emph{SWIRE-selected sample -} The majority of sources in the multiwavelength catalogue were not individually catalogued at 150 MHz by \cite{Sabater2021}. Excluding these sources risks biasing the results towards the minority of radio-bright sources. Therefore we also follow the \citetalias{Smith2021} sample definition and select sources with SWIRE $3.6 \mu \mathrm{m}$ flux density exceeding 10\,$\mu$Jy. This roughly corresponds to the $5\sigma$ detection threshold in the SWIRE $3.6 \mu \mathrm{m}$ data and is in line with SWIRE webpage\footnote{\url{https://irsa.ipac.caltech.edu/data/SPITZER/SWIRE/}} recommendations. \citetalias{Smith2021} adopt the SWIRE $3.6 \mu \mathrm{m}$ selection as these data are extremely sensitive and are less susceptible to the influence of dust obscuration and sample biases than samples identified at shorter wavelengths. We restrict this sample to $z<1$, where \cite{Duncan2021} photo-$z$ estimates are the most reliable, bringing the sample size to 145,162 sources. Finally, we exclude sources at $z<0.05$ to avoid potential IRAC sources with extended radio emission larger than the LoTSS beam size of 6 arcsec. More than 99.5 per cent of sources at $z < 1$ and over 98 per cent of sources at $z < 0.2$ have an FWHM smaller than 6 arcsec \citepalias[see][]{Smith2021}, therefore this redshift cut does not significantly impact our work. The resulting source sample contains 133,544 sources and is referred to as the ``SWIRE-selected sample''.
\end{enumerate}

We use the \cite{Sabater2021} flux densities and uncertainties for the 11 per cent of the SWIRE-selected sample that have 150 MHz counterparts in the \citet{Kondapally2021} catalogue. For the remaining sources, we follow \citetalias{Smith2021} and use the flux densities measured at the 150 MHz image pixel coordinates corresponding to each source. For unresolved sources, these values represent the maximum likelihood estimate of the integrated flux density. Figure\,\ref{fig:rad_lumdist} illustrates the different parameter spaces occupied by the two source samples, with the Radio-selected sample shown in purple, and the SWIRE-selected sample shown in light blue. The marginal histogram to the right of Figure\,\ref{fig:rad_lumdist} indicates the sample size as a function of 150\,MHz luminosity, with the 150\,MHz non-detected sources assigned to the (numerically dominant) lowest luminosity bin if their maximum-likelihood luminosity estimates were lower than this. 
The individually undetected sources (with S/N $< 5$, represented by the blue data points below the lower limit of the purple distribution in Figure\,\ref{fig:rad_lumdist}) are numerically dominant and together can provide significant analytical insights, making it critical to include these sources in our analysis.

\section{SED fitting} 
\label{section:sedfitting}

\begin{table*}
{\footnotesize
\begin{tabular}{lrll}
\toprule
                                  & Parameter                           & Description                                            & Prior/Value                                                            \\
\midrule
Stellar Mass                      & $\log_{10} (M_\star/M_\odot)$                       & Total stellar mass formed $\dagger$            & Uniform; between $6.5$ to $13.5$              \vspace{0.1cm} \\ 
\multirow{4}{*}{Dust attenuation} & $\tau_{\lambda,2}$                  & Diffuse dust V-band optical depth                      & Uniform; between $-1$ and $4$                                    \\
                                  & $n$                                 & Diffuse dust attenuation index                         & Uniform; between $-2$ and $0.5$                                  \\
                                  & $\tau_{\lambda,1}/\tau_{\lambda,2}$ & The ratio of the optical depth of the birth cloud      & Clipped normal; mean = 1, variance = \\
                                  &                                     & to the diffuse dust screen optical depth               & 0.3, between 0 and 2                                                                   \vspace{0.1cm} \\ 
\multirow{5}{*}{Dust emission}    & $U_\text{min}$                      & The minimum intensity of starlight which heats         & Uniform; between $0.1$ and $25$                                  \\
                                  &                                     & up the diffuse dust and ISM                            &                                                                  \\
                                  & $\gamma_e$                          & Fraction of total dust mass affected by $U_\text{min}$ & Log uniform; between $0.001$ and $0.15$                          \\
                                  & $Q_\text{PAH}$                      & The fraction of total dust mass coming from polycyclic & Uniform; between $0.1$ and $10$                                  \\
                                  &                                     & aromatic hydrocarbons (PAHs)                            &                                                                  \vspace{0.1cm} \\ 
\multirow{2}{*}{AGN}              & $f_\text{AGN}$                      & Total luminosity of the AGN defined as a fraction of   & Log uniform; between $10^{-5}$ and $3$                           \\
                                  &                                     & total mid-IR $(4-20 \mu \mathrm{m})$ luminosity                  &                                                                  \\
                                  & $\tau_\text{AGN}$                   & Optical depth of the AGN torus                         & Log uniform; between $5$ and $150$                               \vspace{0.1cm} \\ 
Nebular Emission                  & $\log_{10} (U_\mathrm{gas})$             & Gas-phase ionisation parameter                         & Uniform; between $-4$ and $-1$\\
\midrule
Fixed Parameters                  & IMF                                 & Initial mass function                                  & \cite{Kroupa2001} \\
                                  & $z$                                 & Redshift                                               & Fixed at spec-$z$ if available, else fixed at photo-$z$ \\
                                  & $\log_{10} (Z_\odot)$                    & Stellar metallicity                                    & Fixed at solar.\\
                                  & $\tau_{\lambda,1}$                  & Additional optical depth attenuating the light         & Function of $\tau_{\lambda,1}/\tau_{\lambda,2}$\\
                                  &                                     & from young stars                                       &      \\ 
                                  & $k'(\lambda)$                       & Dust attenuation                                       & \cite{Kriek2013} attenuation curve \\
                                  & $t_\mathrm{bins}$                   & Discrete time binning scheme                           & Fixed, defined in Section\,\ref{section:sedfitting}   \\
                                  & $\log_{10} (Z_\mathrm{gas})$             & Gas-phase metallicity                                  & Equal to $\log (Z_\odot)$ \\
\bottomrule               
\end{tabular}}
\caption{Description of the \pros\,parameters used in the final model. $\dagger$ The \pros\ models use the total stellar mass formed in a source as a free parameter, which is larger than the currently surviving stellar mass. We use the latter for analysis purposes throughout this work.}
\label{table:prospector-model}
\end{table*}

\begin{figure}
  \centering
  \includegraphics[width=\columnwidth]{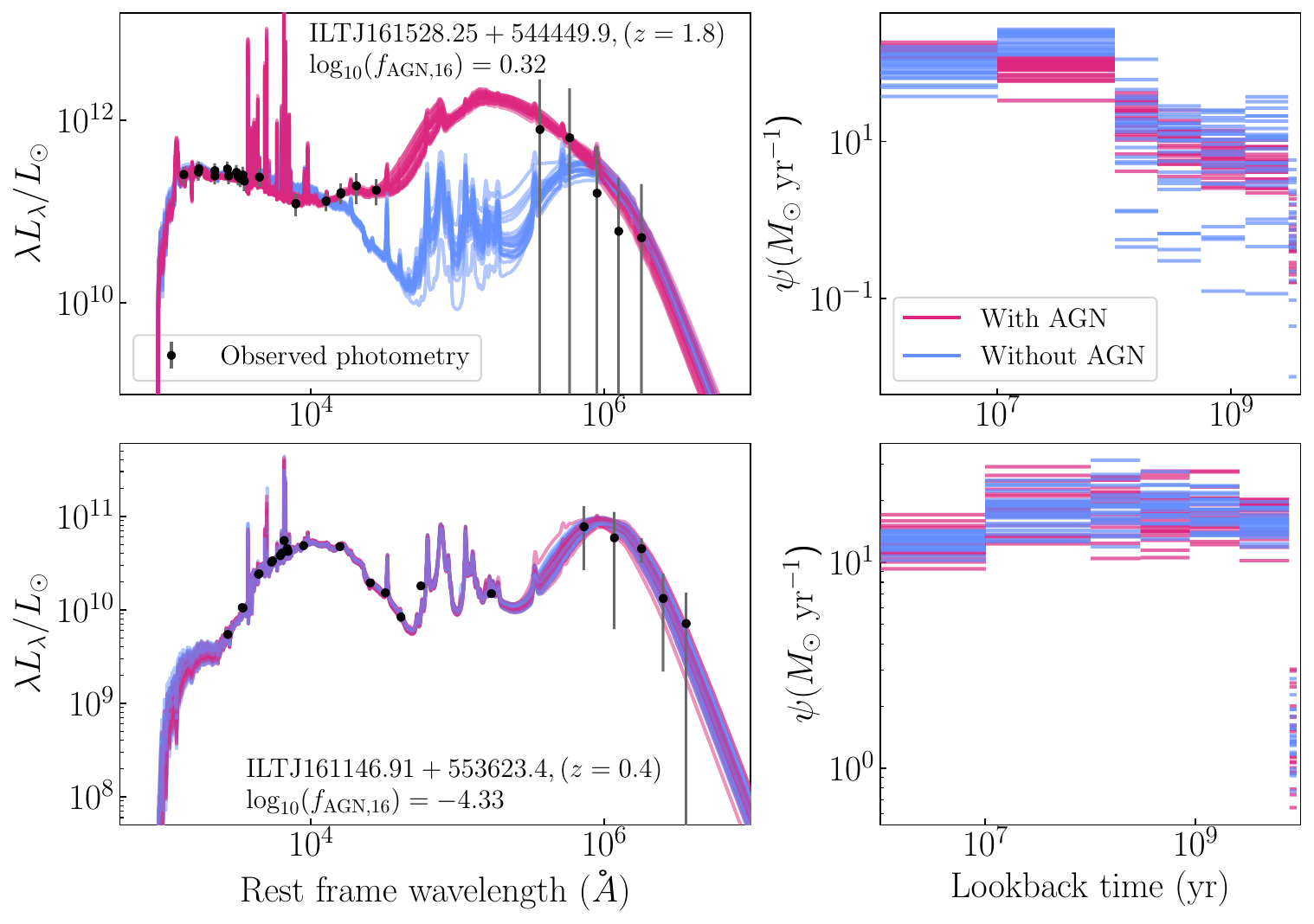}
    \caption{SED fits (left-hand panels) and recovered star formation histories (right-hand panels) for two example radio sources. ILTJ161528.25+544449.9, at a redshift of 1.8, with an estimated $\log_{10}(f_\mathrm{AGN,16}) = 0.32$ (indicating the likely presence of radiative-mode AGN), is shown in the top panels. Meanwhile, ILTJ161146.91+553623.4, at $z = 0.4$ with $\log_{10}(f_\mathrm{AGN,16}) = -4.33$ (implying a likely lack of radiative-mode AGN activity), is displayed in the bottom panels. In left-hand panels the photometry is shown with black dots and error bars, while solid lines represent SED fits randomly drawn from the \dynesty\,chains. In all panels, red lines indicate fits with an AGN component included, while blue lines refer to fits without AGN.}
    \label{fig:sample_sedfits}
\end{figure}

\pros\ \citep{Leja2017,Johnson2021} is an SED fitting code that uses the Flexible Stellar Populations Synthesis \citep[\fsps;][]{Conroy2009,Johnson2023fsps} code to model photometric and/or spectroscopic data and infer stellar population properties. It uses the Padova isochrone table \citep{Bertelli1994,Girardi2000,Marigo2008} and the MILES spectral library \citep{Sanchez2006}. The dynamic nested sampling tool \dynesty\ \citep{Speagle2020} is used to estimate Bayesian posteriors and evidences. We adopt the Initial Mass Function (IMF) from \cite{Kroupa2001}. Since \pros\,does not support evolving metallicities \citep[in contrast to, e.g., \prospect;][]{Thorne2022b}, and even non-evolving metallicities remain poorly constrained based on photometry alone \citep[e.g.][]{Smith2018}, we set the stellar metallicity at solar. While \cite{Thorne2022b} have shown that the stellar masses and SFRs are not systematically offset by this choice of metallicity, we intend to revisit this choice in future works once WEAVE-LOFAR spectra are available. 

Source redshifts have been fixed at spec-$z$ for sources where they are available (5.1 per cent of the radio-detected sample and 1.5 per cent of the SWIRE-selected sample). For the remaining sources, they were fixed at the photo-$z$ estimates from \citet{Duncan2021}. Following \citetalias{Best2023}, neither redshift uncertainties, nor those sources lacking redshift information were considered further. Applying the flags recommended in \cite{Kondapally2021}, we ensure that only those sources with the most reliable photo-$z$ are included in our work.\footnote{The outlier fractions -- defined as sources with $|z_\mathrm{spec} - z_\mathrm{photo}|/(1+z_\mathrm{spec}) > 0.15$, see \citet{Duncan2021} -- for the radio-detected and SWIRE-selected sources, calculated from sub-samples with available spectroscopic redshifts, are 4.3 per cent and 3.2 per cent, respectively. These values fall within the range of the outlier fractions as reported by \cite{Duncan2021}: 1.5 to 1.8 per cent for normal galaxies, and from 18 to 22 per cent for sources with identified AGN components.} 

We use the two-component \cite{Charlot2000} dust attenuation model, which consists of separate birth cloud and diffuse dust screens, to account for the differential attenuation of starlight from younger and older stellar populations. We use the \cite{Kriek2013} attenuation curve which is parametrised by the diffuse dust $V$-band optical depth, the diffuse dust attenuation index, and the ratio of the birth cloud optical depth to the diffuse dust screen optical depth, while energy balance is assumed between dust attenuation and emission \citep[see][for further discussions]{Leja2017}. The three parameter \cite{Draine2007} dust emission templates are used to describe the shape of the IR SED. Mid-IR emission from the AGN torus is described by the CLUMPY models \citep{Nenkova2008A,Nenkova2008B}. These models are parametrised by the $4-20 \mu$m AGN luminosity (which is expressed as a fraction of the total stellar bolometric luminosity) and the optical depth of the AGN torus. Nebular lines and continuum emission are calculated through the implementation of the {\tt \textsc{CLOUDY}} photoionization code \citep{Ferland2013}, wherein the ionizing sources are represented by the stellar populations generated using \fsps\ \citep{Byler2017}. This is parametrised by the intensity of the ionizing spectrum, which is denoted by a dimensionless ionization parameter (and kept free in the model), and the gas-phase metallicity. While \cite{Leja2017} demonstrated that gas-phase metallicity is not adequately constrained for individual galaxies using photometry, works such as \citet{Bellstedt2020}, \citet{Bellstedt2021} and \citet{Thorne2022} have highlighted the benefits of modelling galaxies using an evolving metallicity (e.g. for recovering the best estimates of the cosmic star formation history, the CSFH). Since that functionality is not available in \pros, we have made the pragmatic decision to set the gas-phase metallicity equal to the solar metallicity.

Star formation histories have traditionally been modelled with a number of functional forms \citep[such as constant or exponentially-declining SFHs, commonly referred to as `parametric' SFHs; e.g.][and references therein]{Buat2008,Maraston2010,Gladders2013,Simha2014,Carnall2018,Carnall2019}. Parametric SFHs often struggle to model complex SFH features, such as episodes of star-burst activity, sudden quenching, and rejuvenation \citep{Simha2014,Smith2015,Carnall2019,Leja2019B}. Non-parametric SFHs promise a way out by {\it not} assuming a specific functional form for the SFH. In its simplest form, the lifetime of the galaxy is divided into discrete time bins and the stellar mass formed in each bin is fitted \citep[e.g.,][]{Fernandes2005,Ocvirk2006,Tojeiro2007,Leja2017}. \citet{Iyer2017}, \citet{Leja2019B}, \cite{Lower2020}, and \citet{Ciesla2023} have shown that non-parametric SFH models provide the flexibility to describe the full extent of complex SFHs, potentially leading to more robust estimates of galaxy properties and outperforming parametric SFH models, usually at the expense of a few extra free parameters. Given the inability of photometric data to fully constrain SFHs of galaxies \citep[e.g.][]{Smith2015} the choice of prior distribution plays an important role \citep[e.g.][]{Leja2019B}. 

In this work, we adopt the \pros\ non-parametric ``continuity'' prior, which directly fits for $\Delta\log_{10}(\mathrm{SFR})$ between neighbouring time bins using a Student's $t$-distribution with scale factor $\sigma = 0.3$ and two degrees of freedom $(\nu = 2)$. This prior discourages abrupt changes in SFR between adjacent bins, but remains flexible enough to fit starbursts, star-forming and quenched galaxies \citep[Das et al. \textit{in preparation}]{Leja2019B,Johnson2021,Haskell2023B}. We divide the SFH up such that the first two and the last time bins are kept the same for all sources, covering $0 < t_l < 10$ Myrs, $10 < t_l < 100$ Myrs, and $0.85t_z < t_l < t_z$, respectively. Here, $t_z$ represents the age of the Universe at the object's redshift, and $t_l$ the lookback time. The remaining period between 100 Myrs and $0.85t_z$ is evenly spaced in logarithmic time. The number of bins is selected such that $\log_{10}(\Delta t_l/\mathrm{GYr}) > 0.02$ for each bin. The number of bins ranges from nine (for the lowest redshift sources) to six (for the highest). Following \cite{Smith2012}, an SED fit is deemed acceptable if the best-fit $\chi^2$ is below the 99 per cent confidence threshold for the given number of photometric bands\footnote{These limits were derived based on Monte Carlo simulations fit with the \magp\ model, and it is unlikely to be strictly applicable for our \pros\ results; we intend to test how well this works in a future investigation.}.

Tests show that with this setup, \pros\ produces good agreement with the spectroscopic SFRs (obtained from dust extinction and aperture corrected H$\alpha$ emission line measurements) for the small minority of sources in the MPA-JHU catalogue \citep[][see Appendix \ref{appendix:sdsssfr}]{Brinchmann2004}, and a plausible distribution of derived galaxy properties in the SFR-stellar mass plane relative to the so-called galaxy ``main sequence'' \citetext{e.g. the presence of a main-sequence and distinct clumps of starbursts and quiescent galaxies, as expected from works such as \citealp{Noeske2007} and \citealp{Schreiber2015}}, which is shown in Appendix \ref{appendix:mainseq}. We intend to further investigate the agreement between \pros\ and spectroscopic SFRs in a future work. Table\,\ref{table:prospector-model} summarises the key \pros\,model parameters and the prior distributions that we adopt for this work. 

In Figure \ref{fig:sample_sedfits}, we present the SED output and the recovered SFH for two radio-selected galaxies: one exhibiting strong MIR emission, due to the presence of a radiative-mode AGN, and another without such emission. Both of these sources were fit twice - once with an AGN component included in the fitting process, and once with any possible AGN contribution neglected. It is evident that the source with a strong MIR emission is better fitted when an AGN template is included in the model, however the impact of the inclusion of AGN on the recovered SFHs is more complex (visible in the top right panel of Figure \ref{fig:sample_sedfits}). On the other hand, for the source without an obvious MIR excess, the SED fits as well as the recovered SFH from the two sets of fits are virtually indistinguishable whether an AGN component is included or not. 

\subsection{Comparison of physical parameters} \label{section:estimation_sfr_mass}
The principal physical quantities of interest for this work are the current stellar mass, the SFR and specific SFR (each averaged over the last 10 and 100 Myrs), and the dust luminosity computed between $8-1000\mu$m. The current stellar mass refers to the surviving stellar mass after accounting for losses during evolution (e.g. due to AGB winds, supernovae) and is different from the total stellar mass {\it formed} which is a free parameter in \pros\,models. Throughout this work, the term ``stellar mass'' will refer to the current stellar mass, unless specified otherwise. For each of these parameters, we construct marginalised PDFs by weighting the output \dynesty\,samples (analogous to Markov chains) by their corresponding importance weights. The median likelihood estimates and the 16\tsu{th} and 84\tsu{th} percentile bounds are computed from these PDFs. We compare these parameters to those obtained using the popular SED fitting code \magp, for sources in the radio selected sample for which both \magp\,and \pros\,produced fits that are considered acceptable following the $\chi^2$ based criterion outlined in \cite{Smith2012}. As noted above, the \magp\ SED fitting of LoTSS Deep Fields galaxies was done by \citetalias{Smith2021} and \citetalias{Best2023}, and we refer the reader to these for a detailed account of the process, highlighting here some key differences. For the stellar SEDs \magp\ uses a library of models based on 50,000 exponentially declining SFHs with random bursts superposed, assuming the \cite{Chabrier2003} IMF. It uses the two-component \cite{Charlot2000} dust model and the principle of energy balance to model the dust emission, neglecting any possible contribution of AGN. 

\begin{figure*}
  \centering
  \includegraphics[width=1.9\columnwidth]{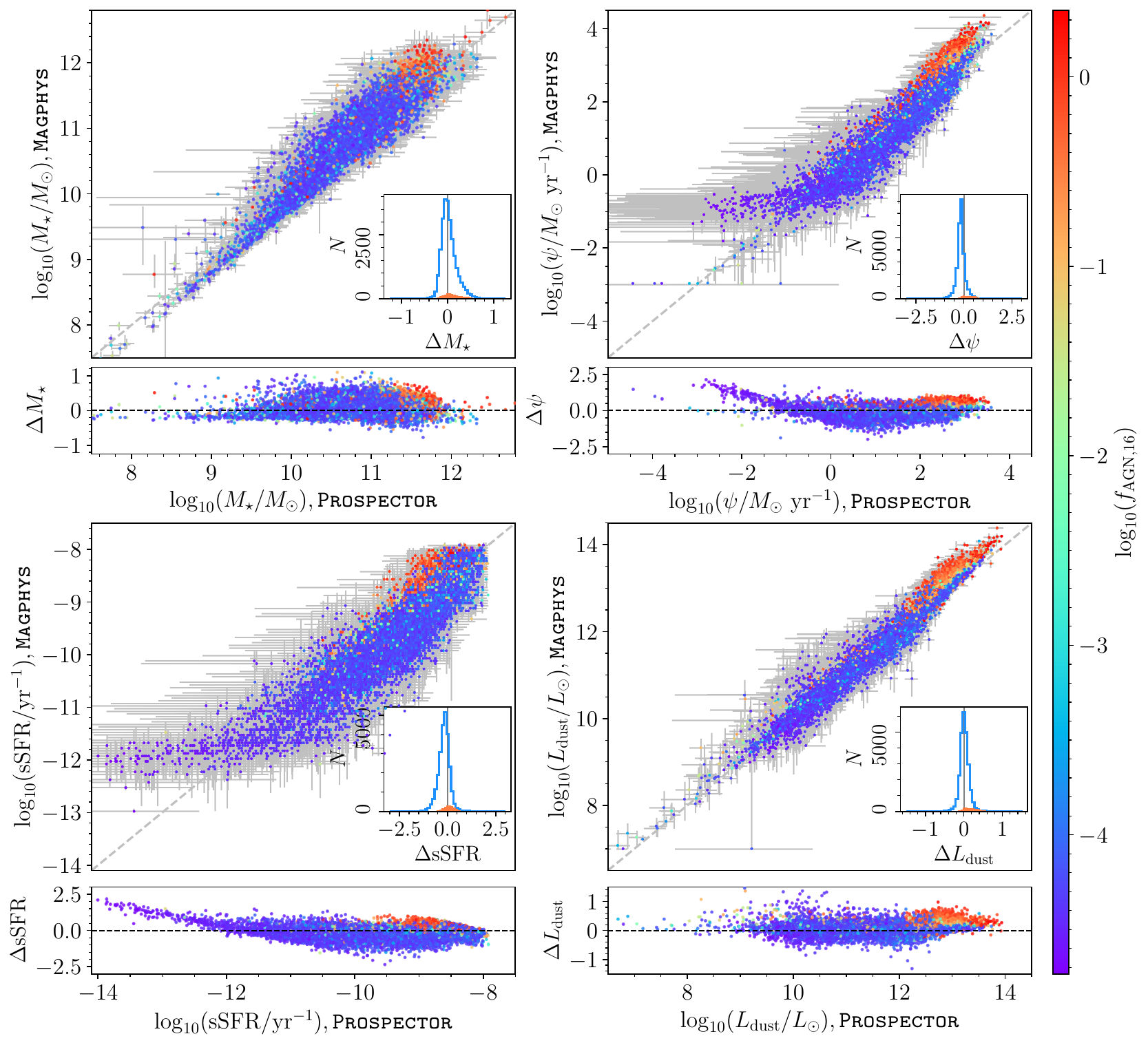}
    \caption{Median likelihood estimates of stellar mass (top-left), star formation rate averaged over the last 100 Myrs (top-right), specific star formation rate averaged over the last 100 Myrs (bottom left), and dust luminosity (bottom-right) with associated uncertainties from \magp\,(plotted along the $y$-axis) compared with those obtained from \pros\,(plotted along the $x$-axis). The sources in each plot are colour coded by the logarithm of the 16\tsu{th} percentile of AGN fraction estimated by \pros. At the bottom of each panel, we plot the difference between the \magp\ and \pros\ estimates ($\Delta X = \log_{10}(X_\mathrm{\magp, median}) - \log_{10}(X_\mathrm{\pros, median})$, where $X$ represents the physical parameter estimates). In the insets we show the distribution of this difference ($\Delta X$), with the blue histogram representing sources with $\log_{10}(f_\mathrm{AGN,16}) < -2$ and the orange filled histogram representing sources with $\log_{10}(f_\mathrm{AGN,16}) > -2$.}
    \label{fig:phys_params_magp}
\end{figure*}

In Figure\,\ref{fig:phys_params_magp}, we compare the median likelihood estimates of stellar mass, SFR (averaged over the last 100 Myrs), sSFR, and dust luminosity estimated using \magp\ and \pros. SFRs have been converted to the \cite{Kroupa2001} IMF where necessary, using the corrections provided in \cite{Madau2014}. These plots are colour-coded by the 16\tsu{th} percentile value of the AGN fraction estimated using \pros, to emphasize the impact of including AGN templates in the models. The ability of the AGN fraction parameter in SED fitting codes to distinguish between AGN dust emission and dust heated by star formation is strongly dependent on the MIR photometric coverage as well as on the S/N ratio of the MIR observations \citep{Leja2018}. Therefore, it is not uncommon for estimated AGN fractions to have large associated uncertainties \citep[e.g.][]{Ciesla2015}. Following \citetalias{Best2023}, we found the 16\tsu{th} percentile of the posterior of the AGN fraction to be a more reliable and robust indicator of AGN activity. Therefore, throughout this work, the term AGN fraction will be used to denote the $16^\mathrm{th}$ percentile values ($f_\mathrm{AGN,16}$). Figure\,\ref{fig:phys_params_magp} shows that estimates from \pros\,and \magp\ are in good general agreement given the uncertainties (sources with high $f_\mathrm{AGN,16}$ aside). For the majority of sources, the stellar masses, SFRs and dust luminosities estimated by the two codes are within 0.3 dex of each other. The \pros\ results show that a large number of sources with high median likelihood stellar masses ($M_\mathrm{star} > 10^{11}\mathrm{~M_\odot}$), SFRs ($> 10^{2}~M_\odot/\mathrm{yr}$), and dust luminosities ($L_\mathrm{dust} > 10^{12}~L_\odot$), also have high AGN fractions, leading to lower \pros\ estimates of stellar mass, SFR, and dust luminosity than \magp\ (which treats stellar emission as the sole origin of emission). Similar trends were observed by \cite{Thorne2022}, who used a method based on \prospect\ AGN fractions\footnote{AGN fraction in \prospect\ is defined as the fraction of the total 5--20 $\mu$m flux contributed by the AGN component.} to identify AGN hosts. They noted a reduction of up to 2 dex in the estimated SFRs for sources with high \prospect\ AGN fractions when modelled with AGN templates, as opposed to when these sources were fitted without AGN templates. Additionally, \pros\ returns lower median likelihood SFRs with larger error bars than \magp\ for galaxies with low SFRs (less than 1 $M_\odot~\mathrm{yr}^{-1}$). While it may seem tempting to consider these larger error bars as a weakness of using \pros\ with non-parametric SFHs, they are in fact one its major strengths, since previous works \citep[e.g.][]{Pacifici2023,Haskell2023A} have shown that \magp\ uncertainties are likely to be underestimated for such galaxies. 

\begin{figure*}
  \centering
  \includegraphics[width=1.9\columnwidth]{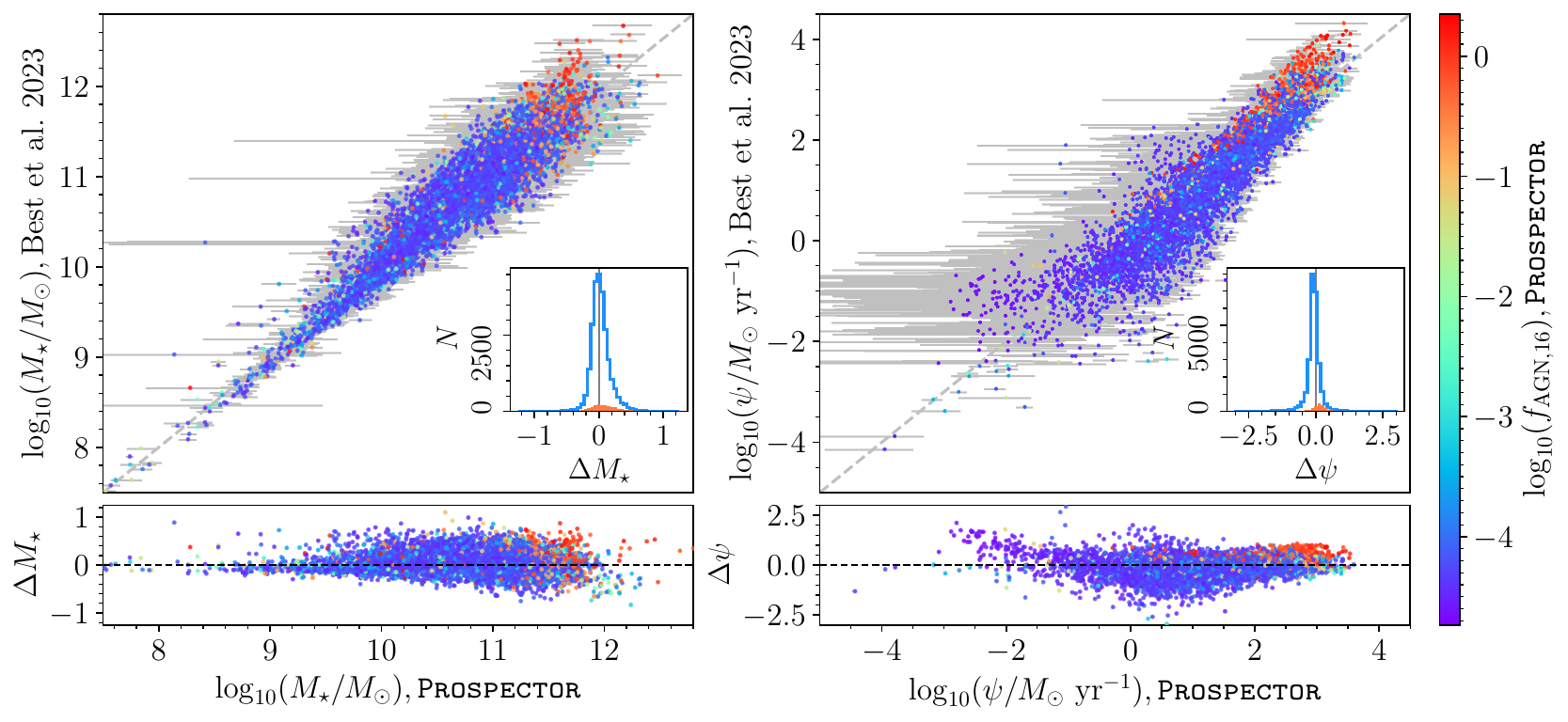}
    \caption{Comparison of the median likelihood estimates of stellar mass (left-hand panel) and the average star formation rate in the last 100 Myrs (right-hand panel) from \pros\,with the consensus estimates from \citetalias{Best2023}. The error bars along the $x$-axis represent the uncertainties in the \pros\ estimates; however, uncertainties are not available for the consensus estimates from \citetalias{Best2023}. The difference between the \citetalias{Best2023} consensus and \pros\ estimates (as the $\Delta$ parameter, see Figure \ref{fig:phys_params_magp}) are plotted at the bottom of each panel. Histograms depicting the distribution of the $\Delta$ parameters are shown in the insets. The blue histograms represent sources with $\log_{10}(f_\mathrm{AGN,16}) < -2$, while the orange filled histograms represent sources with $\log_{10}(f_\mathrm{AGN,16}) > -2$.}
    \label{fig:phys_param_compare_hist}
\end{figure*}

To determine the degree of consistency between our work and \citetalias{Best2023}, in Figure\,\ref{fig:phys_param_compare_hist} we also compare the median likelihood stellar mass and SFR estimates with the ``consensus'' measurements obtained by \citetalias{Best2023} (derived for each galaxy by combining the results from four SED fitting codes, as discussed in Section\,\ref{section:intro}). We find remarkable agreement, similar to the comparison with \magp\ estimates in Figure\,\ref{fig:phys_params_magp}, perhaps unsurprisingly since the \magp\ estimates (or a median of the \magp\,and \bagp\,estimates) are adopted for the large majority of sources not classified as containing AGN by \citetalias{Best2023}. Nevertheless, the good agreement for the high-stellar mass and high-SFR sources where AGN are more prevalent is very encouraging. The principal apparent disagreement between the \citetalias{Best2023} consensus values and the \pros\ estimates appears in the sources with $\log_{10} (\psi_\mathrm{B23}/M_\odot\,\mathrm{yr}^{-1}) \lesssim 0$, however this `difference' is not significant once the error bars are considered. Additionally, \pros\ SED fits indicate that a lower number of radio-detected sources have extremely high SFRs (149 sources with SFR $> 10^{3}~M_\odot~\mathrm{yr}^{-1}$) when compared to both \magp\ SED fits (684) and \citetalias{Best2023} consensus estimates (418). This potentially relieves the tension between the SFR function as derived in works such as \cite{Smit2012}, \cite{Duncan2014} and \cite{Katsianis2017}, and that calculated in recent studies using SFRs estimated from SED fits \citep[which predict an excess of extremely star forming galaxies, e.g.][]{Gao2021}.

\section{Radio source classification}  \label{section:galaxy_classification}
In this section, we look into the classification of radio detected sources into the underlying SFG, RQ-AGN, LERG, and HERG population classes. This involves two distinct parts: (i) identifying those sources hosting a radiative-mode AGN (Section \ref{section:radiative_agn}) and (ii) identifying radio-excess sources (Section \ref{section:radio_excess}). In Section\,\ref{section:final_classifications}, we combine the two criteria to give the overall classification scheme and discuss the characteristics of the various classes.

\subsection{Identification of radiative-mode AGN} \label{section:radiative_agn}

Radiative-mode AGN are typically identified using optical emission-line based diagnostics \citep[e.g.][]{Baldwin1981,Veilleux1987,Kauffmann2003,Best2012,Tanaka2012,Comerford2022} or mid-IR colour-colour cuts \citep[e.g.,][]{Lacy2004,Stern2005,Donley2012,Messias2012}. However, these methods are fraught with several caveats. Observations of the Baldwin-Phillips-Terlevich \citep[BPT;][]{Baldwin1981} emission lines  are strongly affected by dust obscuration as well as by redshift effects. Thus the AGN identification methods that rely on optical emission lines are heavily biased against dust-obscured and low luminosity AGN \citep{Padovani2017,Assef2018}. Furthermore, spectroscopic data are not widely available; \citetalias{Best2023} report that only 5.1 per cent of the radio-detected ELAIS-N1 deep field sources have associated spectroscopic information, though the WEAVE-LOFAR survey \citep{Smith2016} will soon offer complete spectroscopic coverage of these sources\footnote{We note that recent Dark Energy Spectroscopic Instrument release \citep[DESI;][]{DESI2016,DESI2022} has already improved this situation by obtaining spectra for $\sim$25 per cent of the ELAIS-N1 radio sources, and these data are being investigated by Arnaudova et al. ({\it in preparation}).}. Irrespective, selection criteria based on mid-IR colour-cuts often suffer from low signal-to-noise measurements and sample incompleteness, especially when dealing with fainter galaxies. Source samples classified using these broad colour-colour cuts are also likely to be contaminated by a large number of higher-redshift inactive galaxies \citep{Barmby2006,Georgantopoulos2008,Gurkan2014,Donley2012,Messias2012}.

\begin{figure}%
    \centering
    \includegraphics[width=\linewidth]{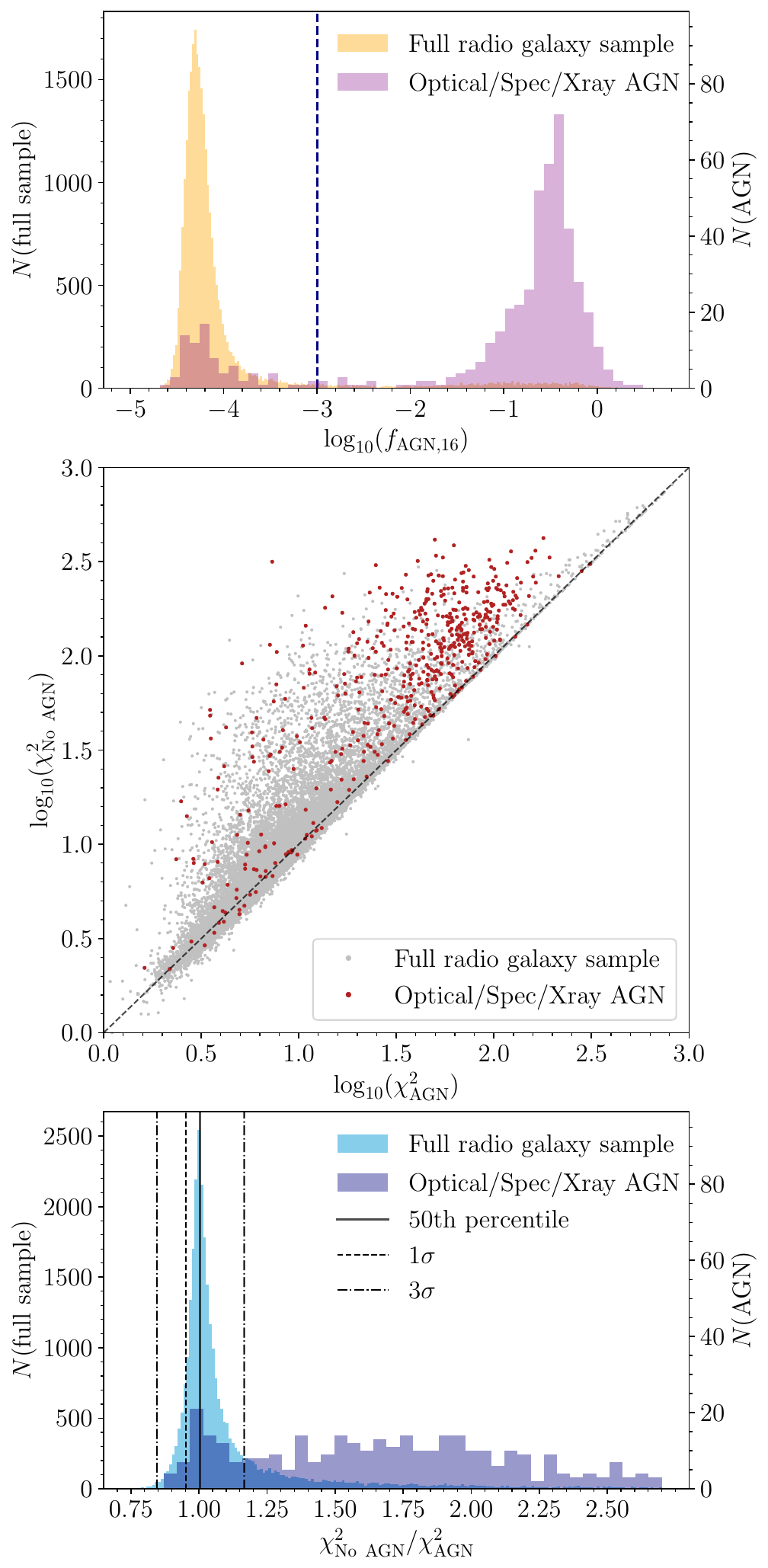}%
    \caption{(Top panel) Histogram showing the distribution of $\log_{10}(f_\mathrm{AGN,16})$. 
    (Middle panel) Comparison of $\chi^2$ arising from \pros\,fits with AGN component included ($\chi^2_\mathrm{AGN}$) along the $x$-axis against the $\chi^2$ from fits without AGN ($\chi^2_\mathrm{No~AGN}$). Sources classified as AGN using previous optical/spectroscopic/X-ray studies are represented as red points. 
    (Bottom panel) Histogram showing the distribution of $\chi^2_\mathrm{No~AGN}/\chi^2_\mathrm{AGN}$. A majority of sources classified as AGN in previous works produce better fits when AGN templates are included in the physical model. 
    }%
    \label{fig:radiative_agn_classification}%
\end{figure}

SED fitting can account for the possible contribution of an AGN whilst simultaneously modelling the host galaxy stars and dust. The AGN fraction parameter in \pros\,directly estimates the fractional contribution of AGN to the total mid-IR $(4-20 \mu \mathrm{m})$ luminosity, and SED modelling can be used to identify AGN at lower luminosities than mid-IR colour cuts. For example, \cite{Leja2018} identify twice as many AGN using \pros\ in the \cite{Brown2014} galaxy sample than what was achieved using the \cite{Stern2012} colour based selection criteria. Additionally, \cite{Thorne2022} identified over 91 per cent of the known emission-line AGN in their sample by employing a similar method based on \prospect\,AGN fractions. The top panel of Figure\,\ref{fig:radiative_agn_classification} shows that the distribution of $f_\mathrm{AGN,16}$ for the radio-detected sample has a prominent peak below $10^{-3}$ and a long tail above this value. Roughly 92 per cent of the sources have $f_\mathrm{AGN,16}$ less than $10^{-3}$. The tail of this distribution is primarily composed of sources that were marked as AGN by \cite{Duncan2021} on the basis of previous optical, spectroscopic, or X-ray studies.  For this latter group, the distribution of $f_\mathrm{AGN,16}$  is bimodal, with a minimum falling between $10^{-3}$ and $10^{-2}$. We identify those sources with $f_\mathrm{AGN,16} > 10^{-3}$ as radiative-mode AGN hosts, noting that the total number of SFGs and AGN we classify in this way depends only weakly on the chosen value for $f_\mathrm{AGN}$ within the limits of $10^{-3}$ and $10^{-2}$.

\cite{Leja2018} and \cite{Thorne2022} demonstrated that incorporating AGN emission in the galaxy model produces better fits (i.e., lower $\chi^2$) for sources with a significant AGN fraction compared to fits without AGN. Therefore to further improve our ability to identify sources hosting radiative AGN, we produce \pros\ fits with and without the inclusion of an AGN component in the model. Figure \ref{fig:sample_sedfits} presents an example where a source with excess MIR emission is better fit when AGN templates are included in the model. Inclusion of AGN templates in the model led to improved fits for 66 per cent of the radio-detected sources (middle panel of Figure\,\ref{fig:radiative_agn_classification}), and 96 per cent of the radio-detected sources which were previously classified as AGN by optical, spectroscopic, or X-ray studies were better fitted when AGN templates were incorporated into the fitting process. 

To systematically identify the sources hosting radiative AGN, we consider the distribution of $\chi^2_\mathrm{No~AGN}/\chi^2_\mathrm{AGN}$ (where the two terms denote the goodness of the best fit SED omitting and including AGN, respectively), which is shown in the lower panel of Figure\,\ref{fig:radiative_agn_classification}. For values below unity, this distribution can be reasonably well-fit by a Gaussian with standard deviation $\sigma = 0.057$. The sources that show a significant improvement in fits after inclusion of AGN templates are identified as AGN. A fit is considered to have significantly improved if $\chi^2_\mathrm{No~AGN}$ exceeds $\chi^2_\mathrm{AGN}$ by a factor of 1.17, corresponding approximately to $3\sigma$.

Combining the two criteria, we classify a source as a radiative-mode AGN if either of the following two conditions is met:
\begin{enumerate}
  \item If the fit including AGN in the model is considered acceptable according to the $\chi^2$ criterion described in Section\,\ref{section:sedfitting}, and $f_\mathrm{AGN,16} > 10^{-3}$, or,
  \item If $\chi^2_\mathrm{No~AGN} > 1.17 \chi^2_\mathrm{AGN}$ and $f_\mathrm{AGN,16}>10^{-3}$.
\end{enumerate}

We flag sources that were not fitted by \pros\ (owing to either a lack of redshift information or not satisfying the recommended cuts in the LoTSS Deep fields catalogue, following \citealt{Duncan2021}) as unclassified. Applying the aforementioned criteria, we identify 3,925 ($\pm 58$) out of the 30,470 radio-selected sources as radiative-mode AGN hosts (where the quoted uncertainties have been obtained by using bootstrap resampling from among our catalogue to generate 10,000 realisations, and calculating the median, 16\tsu{th}, and 84\tsu{th} percentile bounds of the resulting PDF). In this way, we have identified 82 per cent of the 503 optically-/spectroscopically-/X-ray -identified AGN and 73 per cent of the 1540 AGN selected through the \cite{Donley2012} colour cuts\footnote{Upon using $f_\mathrm{AGN,16} = 10^{-2}$ as the threshold instead, we identify  3,396 ($\pm 52$)  sources as radiative-mode AGN. We succeed in identifying 80 per cent and 71 per cent of the optically-/spectroscopically-/X-ray -identified and \cite{Donley2012} colour-cut selected AGN, respectively.}. To assess the performance of our classification scheme, we follow the radiative-AGN selection criteria laid out in \citetalias{Best2023} and use the data from their \magp, \bagp, \cig, and \afit\,catalogues to reproduce their classifications. \citetalias{Best2023} identified 3,129 out of the 30,470 sources as radiative-mode AGN, identifying 83 per cent and 75 per cent of the optically/spectroscopically/X-ray-identified and \citet{Donley2012} AGN, respectively. Although we identify 3,925 sources as hosting radiative AGN as opposed to the 3,129 sources identified by \citetalias{Best2023} (i.e., $\sim 25$ per cent more sources), 84 per cent of the radiative-AGN classified by \citetalias{Best2023} are also identified as such by our method\footnote{If we adopt $f_{\mathrm{AGN,\,16}}>10^{-2}$ the corresponding value is 80\,per cent.}. It is clear that the results obtained by our method to identify radiative-mode AGN, using a single SED fitting code, are comparable to those obtained using four SED codes in \citetalias{Best2023}.

\subsection{Identification of radio-excess sources} 
\label{section:radio_excess}

One of the key tools underpinning our ability to produce robust estimates of SFR and classify radio sources using SED fitting is the relationship between SFR and radio emission in SFGs, thought to arise due to the acceleration of cosmic rays by supernovae. This is further underscored by the well-studied far-infrared radio correlation (FIRC) in SFGs, which has been observed to be linear over several orders of magnitude \citep[e.g.][]{vanderKruit1971,deJong1985,Helou1985,Yun2001,Jarvis2010,Bourne2011,Smith2014,Magnelli2015,Delhaize2017,Read2018,Molnar2021,Delvecchio2021,McCheyne2022}. The precise nature of the FIRC, however, depends on a number of factors, such as the balance between the cosmic ray electron (CRE) escape fraction and the optical depth to UV photons \citep[e.g.][]{Lisenfeld1996,Bell2003,Lacki2010A}.

Nevertheless, radio observations provide a potentially highly attractive means of estimating SFRs for galaxies up to high redshifts, and have enabled direct studies of the correlation between SFR and radio luminosity \citep[e.g.][]{Condon1992,Cram1998,Bell2003,Brinchmann2004,Garn2009,Kennicutt2009,Davies2017,Tabatabaei2017}. Free-free emission is a significant contributor to the radio continuum at frequencies in the GHz range and above \citep{Condon1992,Murphy2011}. In contrast, low frequency radio observations are largely unaffected by free-free emission. Therefore radio surveys at low frequencies, such as LoTSS, are ideal for studying the relationship between radio luminosity and SFR.

Works such as \cite{Brown2017,CalistroRivera2017,Gurkan2018,Read2018,Wang2019,Smith2021} and \cite{Heesen2022} have used the sensitive LoTSS observations to study the so-called ``SFR--$L_\mathrm{150~MHz}$ relation''. At 150 MHz, the primary source of radio emission in galaxies is the non-thermal synchrotron radiation arising either due to stellar processes or accretion activities within the central engine, including jets \citep{Padovani2017}. The latter are commonly exhibited by radio-loud active galactic nuclei \citep[RL AGN; e.g.][]{Urry1995,Wilson1995,Best2005,Harrison2014,Heckman2014,Hardcastle2019,Das2021}. RL AGN can be identified as those sources with a radio luminosity that significantly exceeds what is expected from stellar origin alone \citep{Hardcastle2016,Williams2018,Hardcastle2019,Smolcic2017B,CalistroRivera2017}. The SFR--$L_\mathrm{150~MHz}$ relation can be used to predict the radio emission expected from star formation processes alone, if the SFRs are known. In this section, we use the method of \citetalias{Smith2021} to determine the SFR--$L_\mathrm{150~MHz}$ relation for our SWIRE-selected galaxy sample\footnote{\citetalias{Best2023} used a complementary approach to find the SFR--$L_\mathrm{150\,MHz}$ relation, based on a ``ridgeline'' approach, giving very similar results.}. The main steps of this approach are summarised below, though we refer the reader to \citetalias{Smith2021} for a full discussion of the method. 

It is crucial to account for the 150\,MHz properties of the sources irrespective of whether or not they are detected in the LoTSS catalogue. Failing to do so risks biasing the results towards the radio-bright sources. We therefore use the \pros\,SED fits for the SWIRE-selected sources. This sample, as demonstrated in Section \ref{section:datasets}, includes sources that were not individually catalogued at 150 MHz by \cite{Sabater2021}. It is therefore key to getting a true measure of the low frequency radio source population, given that the source counts of radio sources at low frequencies are dominated by low luminosity SFGs \citep[e.g.][]{Wilman2008,DeZotti2010,Williams2016,CalistroRivera2017,Williams2019,Mandal2021,Hale2023}. Excluding objects flagged as AGN by \cite{Duncan2021} and avoiding those for which \pros\,fails to produce acceptable fits, we are left with 120,307 SWIRE-selected galaxies for our analysis. We first obtain the median likelihood SFRs (averaged over 100\,Myr) along with asymmetric uncertainties derived using the 16\tsu{th}, 50\tsu{th} and 84\tsu{th} percentiles of the derived SFR PDFs, which are equivalent to median $\pm 1\sigma$ in the limit of normally distributed data. Next, we use 150 MHz flux densities and uncertainties from the \citet{Sabater2021} catalogue where they exist, and supplement these for the remaining sources with flux densities (and associated RMS uncertainties) measured from the pixel corresponding to the source position in the ELAIS-N1 maps from LoTSS Deep fields DR1. For each source, we then construct a two-dimensional PDF in the SFR--$L_\mathrm{150~MHz}$ parameter space with logarithmic axes: 70 equally spaced logarithmic SFR bins between $-3 < \log_{10}(\psi/M_\odot~\mathrm{yr}^{-1}) < 3$ and 180 equally spaced log-bins between $17 < \log_{10}(L_\mathrm{150~MHz}/\mathrm{W~Hz}^{-1}) < 26$. This PDF is populated for each source by creating 100 random samples from a normal distribution based on the median-likelihood \pros\ SFR estimates, accounting for the asymmetric uncertainties. 

For the 150\,MHz luminosities we use 100 independent samples from a symmetric normal distribution in linear space, centred on the luminosity obtained from the flux density estimate (either from the catalogue, or from the pixel measurements) assuming the best redshift and a spectral index of $\alpha=0.7$. The standard deviation on the luminosity distribution is obtained by scaling the uncertainties on the flux density measurements such that the signal-to-noise on the derived luminosity is equal to that on the measured flux densities. The 2D PDFs for each source are then summed, to generate a PDF that accounts for the uncertainties in both the SFR and $L_\mathrm{150~MHz}$. To ensure that sources with low S/N and negative radio luminosities were included (since not doing so would censor the distribution and lead to skewed median-likelihood luminosity measurements), we follow \citetalias{Smith2021} and arbitrarily assign samples with $\log_{10}(L_\mathrm{150~MHz}) <17$ to the lowest $L_\mathrm{150~MHz}$ such that each source is equally represented in the 2D stacked PDF. Similarly, we arbitrarily assign samples with $\log_{10}(\psi/M_\odot~\mathrm{yr}^{-1}) < -3$ to the lowest log SFR bin. We then calculate the 16\tsu{th}, 50\tsu{th}, and 84\tsu{th} percentiles of the $L_\mathrm{150~MHz}$ distribution in each log SFR bin, with the 50\tsu{th} percentile values denoting the median likelihood SFR--$L_\mathrm{150~MHz}$ relation. 

By fitting a straight line to the stellar mass-independent relation given by \citetalias{Gurkan2018}:
\begin{equation} \label{eq:indep}
  L_\mathrm{150~MHz} = L_1~\psi^\beta,
\end{equation}

\noindent we obtain a best-fit estimate of $\beta = 1.019\pm0.009$ and $\log_{10}L_1 = 22.024\pm0.006$. Uncertainties are computed using \emcee, which was run with 64 walkers and a chain length of 10,000 samples. In comparison, the following best-fit estimates were obtained in past works: $\beta = 1.07\pm0.01$ and $\log_{10}L_1 = 22.06\pm0.01$ (\citetalias{Gurkan2018}); $\beta = 1.058\pm0.007$ and $\log_{10}L_1 = 22.221\pm0.008$ (\citetalias{Smith2021}); $\beta = 1.08\pm0.06$ and $\log_{10}L_1 = 22.24\pm0.07$ (\citetalias{Best2023}). Unlike the studies by \citetalias{Gurkan2018} and \citetalias{Smith2021}, our best fit slope is consistent with unity after factoring in uncertainties. 

\begin{figure}%
    \centering
    \includegraphics[width=\linewidth]{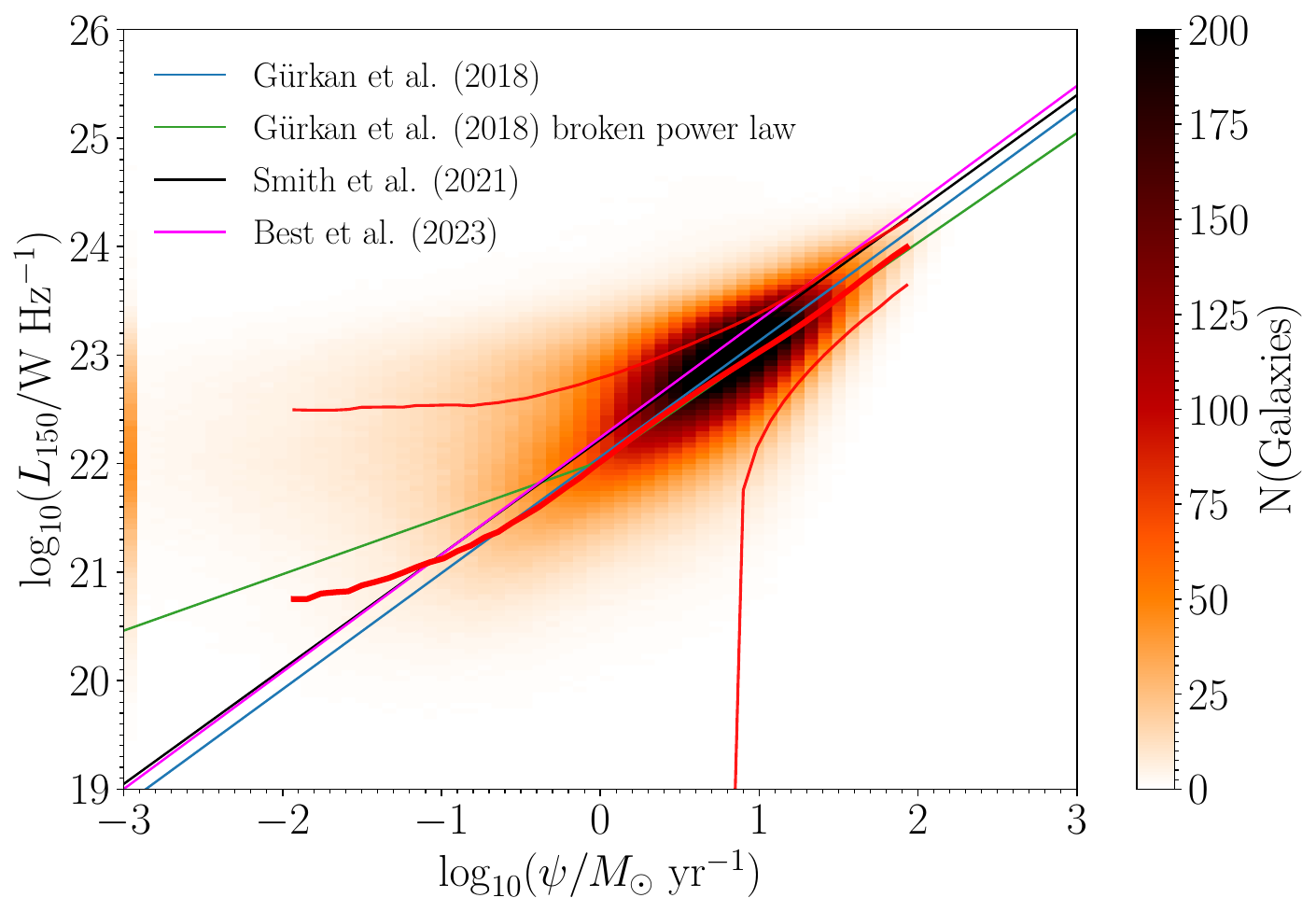}%
    \caption{Stacked two dimensional heatmap showing the SFR--$L_\mathrm{150~MHz}$ plane, constructed by summing over the PDFs of sources in the SWIRE-selected sample. The thick red curve represents the median likelihood SFR--$L_\mathrm{150~MHz}$ relation. The accompanying thin red curves indicate the 16\tsu{th} and 84\tsu{th} percentiles of the $L_\mathrm{150\,MHz}$ distribution at each SFR, respectively. The blue and green solid lines represent the single slope and broken power law relations from \citetalias{Gurkan2018}. The black and magenta solid lines represent the best fit relations from \citetalias{Smith2021} and \citetalias{Best2023} respectively. }
    \label{fig:pros_l150sfr_massindep}%
\end{figure}

As discussed at length in \citetalias{Smith2021}, the SFR--$L_\mathrm{150~MHz}$ relation has significant dispersion, which we must account for in order to identify those sources with a radio excess. To do this, we consider the difference between the 50\tsu{th} and 84\tsu{th} percentiles\footnote{We have elected to use the 50\tsu{th} and 84\tsu{th} percentiles rather than the 16\tsu{th} and 50\tsu{th} percentiles since the distribution is asymmetric, with increased scatter on the bottom side due to the lower signal to noise ratio of the fainter sources in the 150\,MHz maps.} of the $L_\mathrm{150\,MHz}$ distribution for $\log_{10} (\psi/M_\odot~\mathrm{yr}^{-1}) > 1.5$, finding $\sigma = 0.289$\,dex. This is in good agreement with the results of \cite{Cochrane2023}, who found a scatter of 0.3 dex by comparing the shapes of SFR functions for radio- and SED-estimated SFRs. We specifically avoid including sources with SFRs less than $1.5 M_\odot \mathrm{yr}^{-1}$ to mitigate the impact of large uncertainties associated with the best-fit relation at low SFRs (e.g. Figure\,\ref{fig:pros_l150sfr_massindep}). 

\begin{figure}%
    \centering
    \includegraphics[width=0.95\columnwidth]{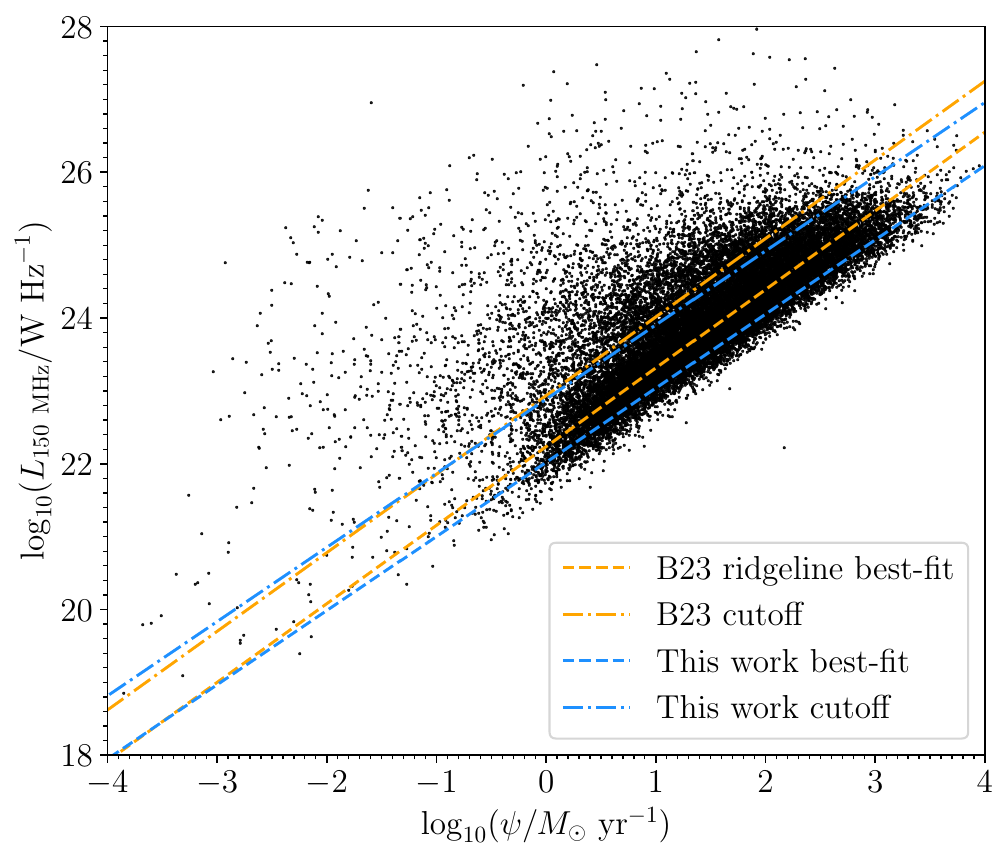}%
    \caption{Maximum likelihood 150 MHz radio luminosities of the radio-detected sample plotted as a function of the \pros\,estimated median SFRs. The best-fit SFR--$L_\mathrm{150~MHz}$ relation (obtained from the SWIRE-detected sample and therefore including radio non detections) and the radio-excess cut-off used in our work (blue) and in \citetalias{Best2023} (orange) are overplotted.}%
    \label{fig:radio_agn_classification}%
\end{figure}

Using this information, we identify sources that exceed the best-fit SFR--$L_\mathrm{150~MHz}$ relation by 0.867 dex (equivalent to a 3$\sigma$ excess) as having {significant} radio-excess (this criterion is indicated by the blue dot-dashed line in Figure\,\ref{fig:radio_agn_classification}). Additionally, as seen in Section \ref{section:estimation_sfr_mass}, SED fitting cannot accurately measure the SFRs for sources with very low SFRs, and thus the radio-excess classifications of these sources are likely to be unreliable. To avoid this issue, \citetalias{Best2023} classified 0.4 per cent of their sources which had SFRs $< 0.01 M_\odot~\mathrm{yr}^{-1}$ as radio-excess only if their radio luminosities exceeded that expected for a source with SFR $ = 0.01 M_\odot~\mathrm{yr}^{-1}$ by 0.7 dex. If the radio luminosity of such a source was below that value but still surpassed the radio-excess threshold determined by the SED-fitted SFR estimate, then it was designated as Unclassifiable. We follow \citetalias{Best2023} and similarly mark 0.1 per cent of our sample as Unclassified. Among the 30,470 sources in the radio-selected sample, we identify 5,250 $\pm$ 66 sources as having a radio excess (\citetalias{Best2023} identified 4,786 sources). Uncertainty in the count of radio excess sources is calculated using bootstrapping, in a similar manner that was adopted for radiative-mode AGN identification in Section \ref{section:radiative_agn}. The median likelihood SFR estimates from \pros\,are lower than the consensus SFRs estimated by \citetalias{Best2023} for 92 per cent of the sources where the radio-excess classifications differ. In these cases, the radio-excess thresholds calculated for the \pros\,results are lower than the thresholds based on \citetalias{Best2023}'s results. We therefore classify about 10 per cent more sources as radio-excess than \citetalias{Best2023}. 

We mark sources with unacceptable fits according to the $\chi^2$ threshold criterion as ``Unclassified'', since the \pros\ model does not apply and we cannot use the SFR estimates to quantify the expected 150\,MHz luminosity. This results in a higher number of sources being flagged as unclassified compared to \citetalias{Best2023}; we have 1,532 compared to 240 (these numbers differ from those shown in Table\,\ref{table:classification} since there the numbers include sources irrespective of whether they have redshifts or are flagged in the cuts recommended by \citealt{Duncan2021}). Our classifications are in agreement with \citetalias{Best2023} for 77 per cent of the sources they classified as radio-excess and 90 per cent of the sources they identified as not having a radio excess. While there are non-negligible areas of implied disagreement among the radio-loud classification (e.g. the $\sim 23$ per cent of the \citetalias{Best2023} radio-excess sources not identified using our method which are not apparent in Figure \ref{fig:radio_agn_classification} since the \citetalias{Best2023} consensus estimates are not plotted), this type of classification is inherently uncertain: we do not know what the ``correct'' answer should be for these sources (e.g. Drake et al. {\it in preparation}). Irrespective, the good general agreement in the radio-AGN classification between the two works offers significant encouragement.

\subsection{Final classifications} \label{section:final_classifications}

Here, following \citetalias{Best2023}, we combine the radiative-mode and radio-excess AGN classifications discussed in the previous sections to assign each source to one of the five classes: SFG, RQ\,AGN, LERG, HERG and Unclassified. Specifically: 

\begin{itemize}
\item sources that have been identified as neither radiative-mode nor having a radio excess are classified as \emph{SFG};
\item those that are identified as radiative-mode AGN hosts, without a radio excess, are classified as \emph{RQ\,AGN};
\item objects not identified as radiative-mode AGN hosts but showing a significant radio excess are classified as \emph{LERG};
\item hosts of radiative-mode AGN with significant radio excess are classified as \emph{HERG}; and finally
\item sources for which we are unable to obtain an acceptable \pros\ SED fit remain \emph{Unclassified}. 
\end{itemize}

\begin{table}
\centering
\begin{tabular}{lccccc}
\hline
             & SFGs   & RQ AGN & LERGs  & HERGs & Unc. \\ \hline
This    & 21257 $\pm$ 83 &  2431 $\pm$ 47  &  4393 $\pm$ 61  &   857 $\pm$ 28 & 2672 $\pm$ 49   \\
work          &    &     &     &     &     \\ 

\citetalias{Best2023}          & 22720  &  2779  &  4287  &   510 & 1314   \\ \hline
\end{tabular}
\caption{The number of star forming galaxies, radio-quiet AGN, low-excitation radio galaxies, and high-excitation galaxies, identified in the ELAIS N1 deep field using our classification scheme, along with uncertainties calculated using bootstrap resampling. \citetalias{Best2023} classification results are also included for comparison.}
\label{table:classification}
\end{table} 

\begin{figure}%
    \centering
    \includegraphics[width=\columnwidth]{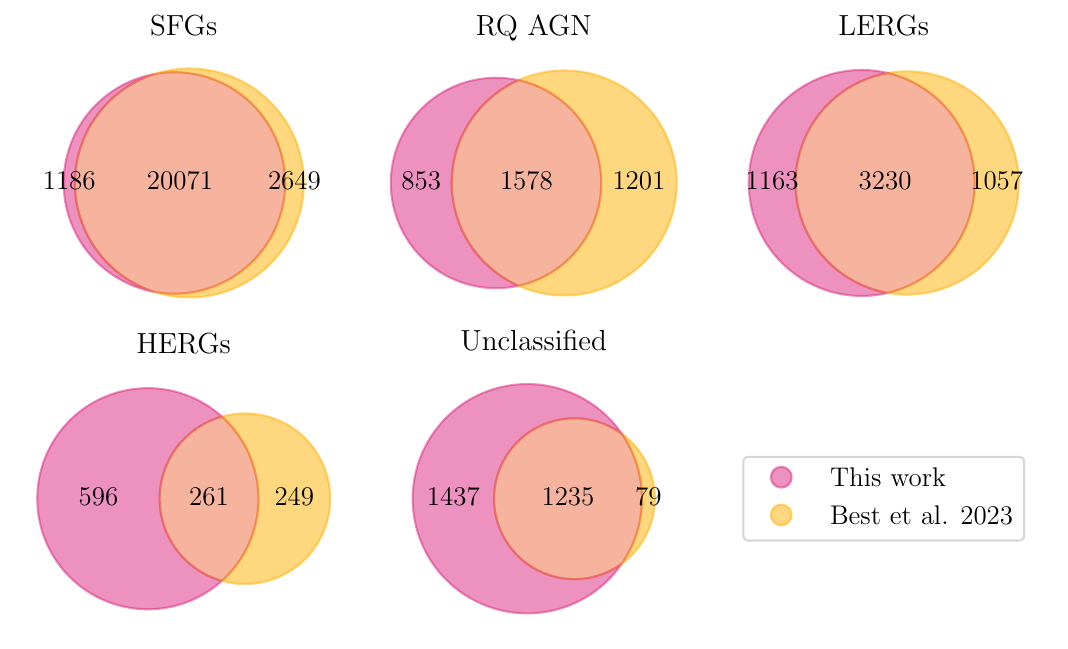}%
    \caption{Venn diagrams showing the number of sources in each class (star-forming galaxies, radio-quiet AGN, low- or high-excitation radio galaxy) that have been identified by our method (pink), by \citetalias{Best2023} (yellow) and by both (overlap region).}%
    \label{fig:classifications_venn}%
\end{figure}

\noindent The number of sources in each category is shown in Table\,\ref{table:classification}, while the degree of overlap between the sources identified following our classification scheme and those presented by \citetalias{Best2023} is illustrated in Figure \ref{fig:classifications_venn}.

Overall, our classification method identifies similar numbers of SFGs (6 per cent lower), RQ\,AGN (12 per cent lower) and LERGs (2 per cent higher) relative to the results of \citetalias{Best2023}. However, there are larger differences apparent in the number of HERGs and unclassified sources, where our numbers are 68 and 203 per cent larger. The former change -- in the number of HERGs -- results primarily from the details of the precise definition of a radio excess that we discussed in Section\,\ref{section:radio_excess} relative to \citetalias{Best2023} (and shown in Figure\,\ref{fig:radio_agn_classification}). This change represents an increase from 1.6 per cent to 2.7 per cent of the radio-selected catalogue, and though this increase is significant, HERGs remain a small minority of the overall source population, perhaps highlighting the difficulty of this task. In the case of \citetalias{Best2023}, a source is flagged as Unclassified if it fails to produce an acceptable fit in multiple codes. Consequently, they report significantly fewer Unclassified sources compared to our results. While we do not formally classify the sources with unacceptable \pros\,fits, roughly 70 per cent of the sources would be consistent with SFGs, 10 per cent with RQ\,AGN, 13 per cent with LERGs, and 3 per cent with HERGs, if we were to disregard the checks for fit acceptability. Interestingly, this is in good agreement with the final classification results, indicating that the ability of \pros\ to produce an acceptable fit for a source is approximately independent of the underlying population class. This could suggest that the $\chi^2$ based acceptability criterion we have used on the \pros\,SED fits may be conservative, and we intend to investigate this possibility in a future work.

\begin{figure*}
  \centering
    \includegraphics[width=0.97\linewidth]{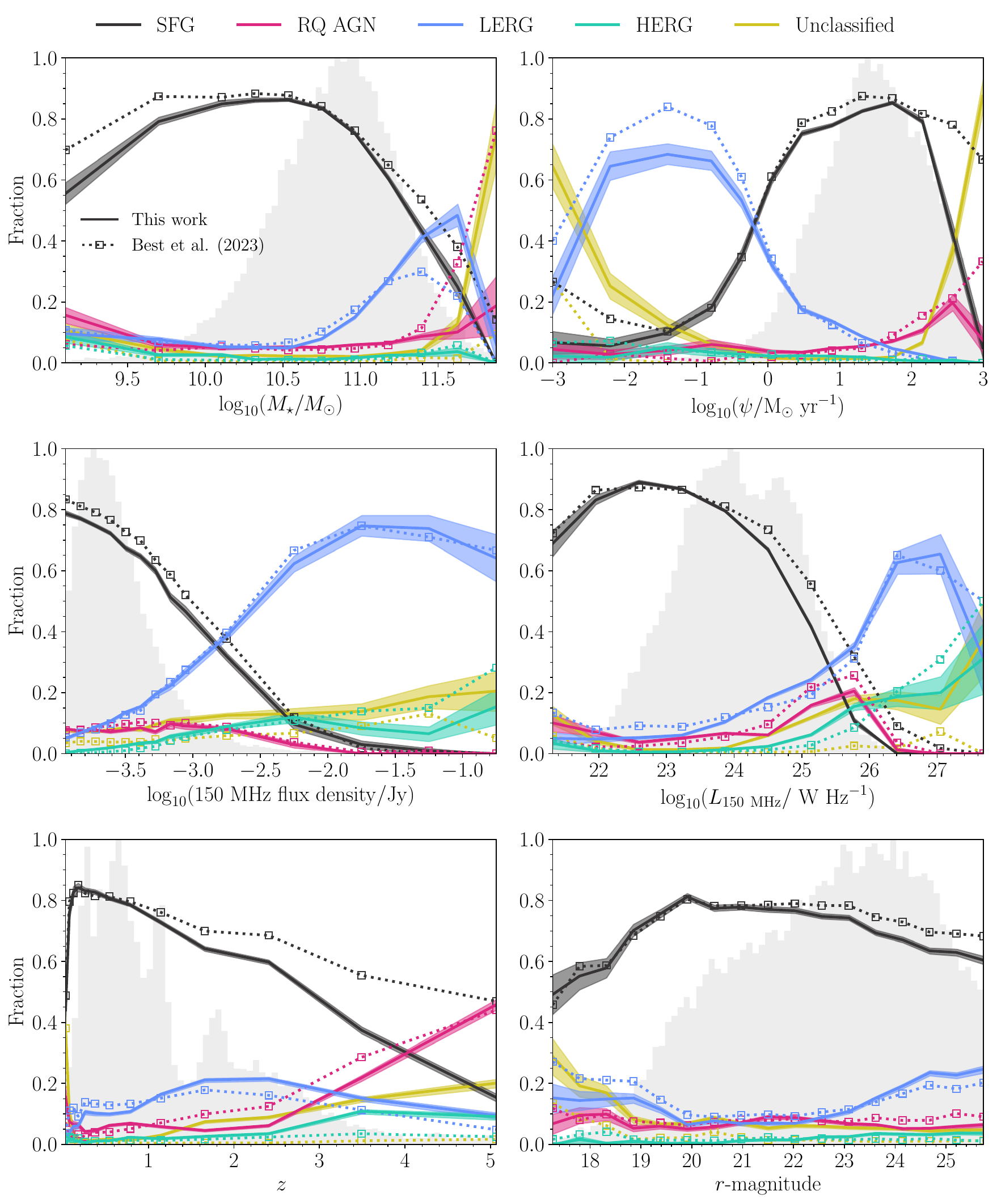} %trim={2cm 0 2cm 2cm},clip,
      \caption{Sources classified as SFG (in black), RQ-AGN (orange), LERG (blue), HERG (purple), and the sources marked unclassified (in green), represented as the fraction of total sources. We show the distribution of sources in each galaxy class as a function of (top-left panel) \pros\,median likelihood stellar mass estimates, (top-right panel) \pros\,median likelihood SFR estimates, (centre-left panel) 150 MHz radio flux density, (centre-right panel) 150 MHz radio luminosity, (bottom-left panel) redshift, and (bottom-right panel) optical $r$-band magnitude. The dotted lines with open squares represent the fractions of sources in each galaxy class from \citetalias{Best2023}. In the top-left and top-right panels, we plot the \citetalias{Best2023} source fractions against the \citetalias{Best2023} consensus stellar mass and SFR estimates, respectively. The distribution of the whole sample as a function of the respective quantities is shown in the background, as a normalised grey histogram with peak being equal to unity.}
      \label{fig:popfractions}
  \end{figure*}

Similarly to \citetalias{Best2023}, we further examine the demographics of the different AGN classes based on stellar mass, SFR, 150 MHz radio flux density, radio luminosity, redshift, and optical $r$-band magnitude; these results are shown in Figure\,\ref{fig:popfractions}. In each of these figures, our results are shown as the solid lines (coloured according to the classifications as indicated in the legend) and with shaded areas according to the uncertainties implied assuming Binomial statistics \citep[e.g.][]{Cameron2011}. The corresponding values from the \citetalias{Best2023} catalogue are shown as squares with dotted lines of the relevant colour, enabling a direct comparison.\footnote{The \citetalias{Best2023} results shown in Figure\,\ref{fig:popfractions} do not exactly match those plotted in \citetalias{Best2023} figure 9, since \citetalias{Best2023} applied additional cuts in some of their plots, e.g. $z<2$ for the 150 MHz radio luminosity plot (P. Best, \textit{private communication}), whereas we use the full sample to directly compare with the published catalogue.}

  \begin{figure}
    \centering
    \includegraphics[width=1\linewidth]{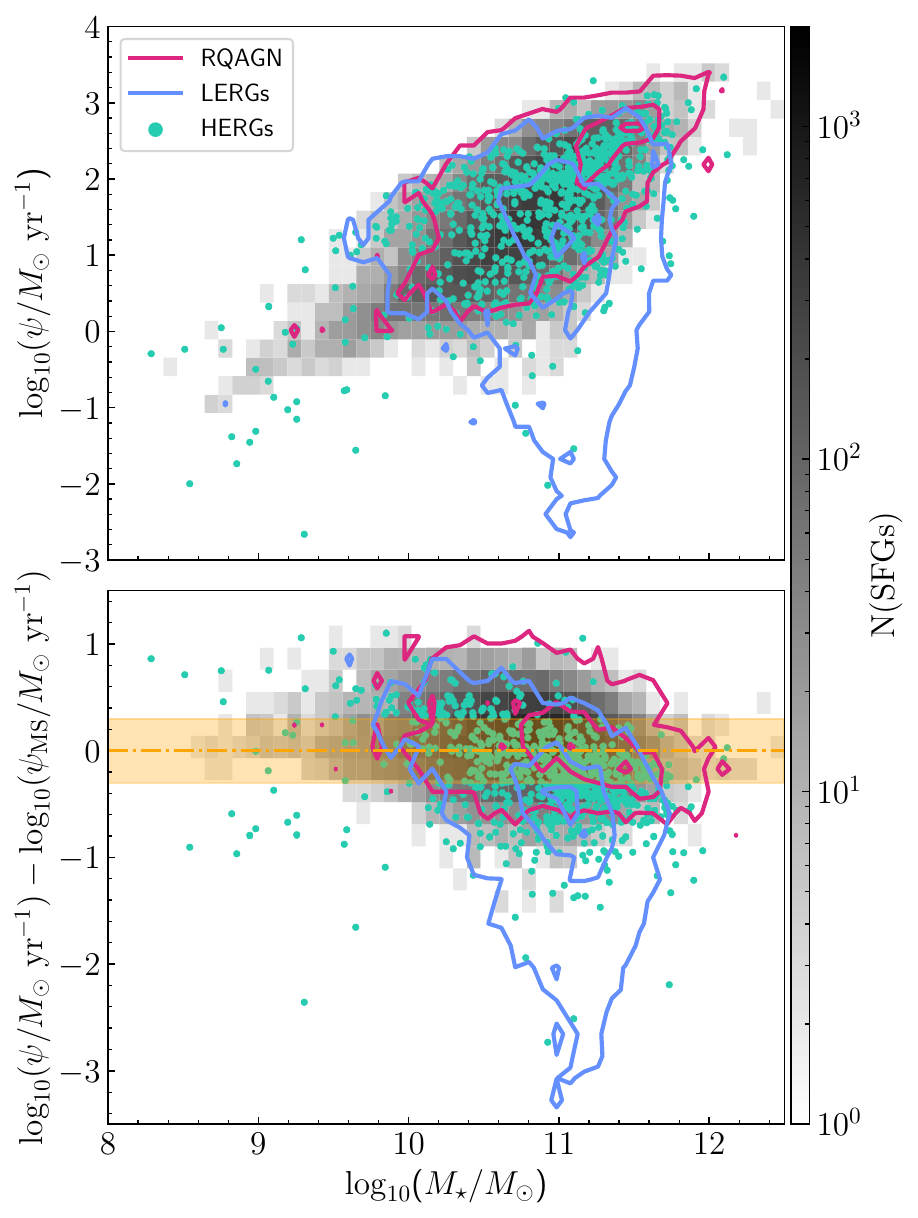}
    \caption{Distribution of the different source classes in the stellar mass--SFR plane: the shaded 2D histogram represents the SFGs, the overlaid purple and blue contours represent the RQ AGN and LERGs, respectively, and the cyan circles denote the HERGs. In the bottom panel, the solid orange line and the shaded orange region represent the \citet{Schreiber2015} parametrisation of the main sequence converted to our adopted \citet{Kroupa2001} IMF with the typical 0.3 dex scatter around it. Contour levels are selected to encompass 5, 50, and 90 per cent of the corresponding sample.}
    \label{fig:mainseq_classi}
\end{figure}

The upper panels of Figure\,\ref{fig:popfractions} show the distribution of the different classes based on the median likelihood estimates of stellar mass and SFR from \pros. SFGs are the dominant population class for most of the low-to-intermediate stellar mass ranges. At higher stellar masses, the number of AGN sees a steep rise, led by LERGs. Additionally, LERGs dominate the source fractions at SFRs below $\sim1~M_\odot\,\mathrm{yr}^{-1}$, supporting the view that they are hosted by massive, quiescent galaxies \citep[e.g.,][]{Tasse2008,Smolcic2009,Best2012,Heckman2014,Williams2018}. Overall, the stellar mass and SFR demographics in Figure\,\ref{fig:popfractions} are similar to those observed by \citetalias{Best2023}, with the principal differences  attributable to the larger number of Unclassified sources in this work (which become the dominant component at the highest stellar masses and SFRs). 

The remaining panels in Figure\,\ref{fig:popfractions} depict the fractions in each class as a function of 150\,MHz flux density, luminosity, redshift and $r$-band magnitude. The demographics of each source type as a function of the 150 MHz radio flux density are generally in agreement with the findings of \citetalias{Best2023}. The LERG and HERG source fractions as functions of 150 MHz radio luminosity follow similar trends as seen in \citetalias{Best2023}. We observe an abundance of LERGs at the highest radio luminosities. It must be noted here that the ELAIS-N1 field may not probe enough volume to reach the very highest radio powers where HERGs dominate. 
Nevertheless, at the very highest radio luminosities, the HERG fraction sees a steep rise, and the fractions of LERGs and HERGs appear to be roughly equal. This steep rise in HERG fraction strongly hints at HERGs becoming the dominant class at higher luminosities, which is consistent with expectations \citep[e.g.,][]{Best2012,Heckman2014,Pracy2016}. 
The population of RQ\,AGN remains relatively stable at around 10 per cent below 1\,mJy and sees a steady decline above that. Our findings are consistent with \citetalias{Best2023}'s results and are lower than the observed fractions of approximately 15-20 per cent reported in studies such as \cite{Simpson2006} and \cite{Smolcic2017B}. 
This difference is likely because of the RQ\,AGN typically having flatter spectral indices than SFGs \citep[e.g.][]{Gloudemans2021}, and therefore as noted by \citetalias{Best2023}, they are likely to be less prominent at the low frequencies probed by LOFAR than at GHz frequencies and higher.
The transition from a sample dominated by star formation to one dominated by radio-loud AGN occurs around 1.6 to 2 mJy, which closely aligns with \citetalias{Best2023}'s finding of 1.5 mJy. As noted by those authors, this is consistent with the transition occurring around $200-250 \mu$Jy at a higher frequency of 1.4\,GHz, assuming typical radio spectral indices \citep{Padovani2016,Smolcic2017B}.

We detect radio sources of each class at all redshifts and optical $r$-band magnitudes. At $z > 1$, we observe lower SFG and RQ-AGN fractions compared to \citetalias{Best2023}, but an increased fraction of LERGs. Our results suggest that SFGs dominate up to redshifts $z \sim 4$, and are subsequently overtaken by RQ\,AGN. We observe a steeper decline in the SFG fraction with redshift compared to \citetalias{Best2023}, whose classifications suggest a transition from a star-formation dominated sample to an AGN dominated sample at $z \sim 4.5$. We note that the reduced SFG fractions relative to \citetalias{Best2023} could be caused by (i) a larger number of sources being classified as radio-loud AGN at high redshifts, owing to \pros\,estimating lower SFRs for sources with extreme SFRs (often found at very high redshifts) than \citetalias{Best2023}, therefore having lower radio-excess thresholds, and (ii) a larger number of potential SFGs being flagged as `Unclassified’ in this work on account of the unacceptable \pros\ SED fit $\chi^2$. The LERG fraction peaks around $z \sim 2$ and subsequently decreases, while RQ-AGN, HERGs and Unclassified sources all increase at $z > 3$. This hints at the existence of an evolutionary relationship between a more energetic, radiative-AGN dominated source population in the early stages of the universe and the jet-mode population at cosmic noon. We draw the attention of the reader to the increase in the fraction of unclassified sources with redshift, with approximately 20 per cent of sources labelled as unclassified at $z = 5$ in our results, and advise caution when using our classifications at high redshifts.

In Figure \ref{fig:mainseq_classi}, we show the SFR vs stellar mass plane, populated by a shaded 2D histogram showing the number of SFGs at each point colour-coded according to the legend on the right-hand side, which is shared between both the top and bottom panels. In the lower panel, the SFRs have been calculated relative to the so-called ``main sequence'' relation from \cite{Schreiber2015}, converted to our adopted \cite{Kroupa2001} IMF. The typical scatter of $\pm 0.3$ dex is indicated by the shaded orange region. Overlaid on the SFGs are contours delineating the locations of the RQ AGN (purple) and LERGs (blue), while the HERGs are represented by cyan circles. The contour levels have been chosen such that they include 5, 50, and 90 per cent of the sample. The SFGs lie broadly around the main sequence as expected, although the observed scatter is slightly greater than the canonical $\pm 0.3$ dex reported in literature \citep[e.g.][]{Tacchella2015}. As seen in the bottom panel, a considerable portion of the SFG distribution lies above the main sequence, indicating the presence of a population of starbursts in the radio-detected sample. The RQ AGN occupy a similar region of parameter space as the SFGs, albeit biased towards the high stellar-mass end of the distribution, consistent with our expectations that AGN predominantly reside in higher-mass galaxies \citetext{e.g., \citetalias{Gurkan2018} and \citealp{Kondapally2022}}. Notably, we find a significant population of LERGs ($\sim33$ per cent) to lie on or above the main sequence, consistent with the distinct populations of star-forming and quiescent LERGs discussed by \citet{Kondapally2022}. At higher stellar masses, LERGs are hosted across all SFRs, whereas at lower stellar masses, only star-forming galaxies host LERGs. This is in agreement with the findings of Kondapally et al. (\textit{in preparation}), and supports their hypothesis that different fuelling mechanisms may be prevalent in different classes of LERGs. Lastly, we observe that 58 per cent of the HERGs lie within the typical 0.3 dex range of the main sequence, with the significant majority of HERGs falling within a 0.6 dex scatter. Our classifications thus align with works such as \citet{Best2012}, \citet{Best2014}, \citet{Pracy2016}, and \citet{Kondapally2022}, who noted that HERGs were typically hosted by star-forming galaxies.

\section{Stellar mass dependence in the SFR--150 MHz radio luminosity relation} \label{section:massdependence}

As discussed in Section \ref{section:radio_excess}, LoTSS data have been extensively used to study the SFR--$L_\mathrm{150~MHz}$ relation. \cite{Brown2017}, \cite{Wang2019}, and \cite{Heesen2022}, found a super-linear relationship (slope greater than unity) between SFR and 150 MHz radio luminosity. \citetalias{Gurkan2018} found an upturn towards larger radio luminosity at low SFRs, which is in disagreement with calorimetric arguments, and in the opposite direction to the downturn reported by \citet{Bell2003}. \citetalias{Gurkan2018}, \cite{Read2018}, and \citetalias{Smith2021} also found evidence for a strong stellar mass dependence in the SFR--$L_\mathrm{150~MHz}$ relation. In this section, we seek to address the question of stellar mass-dependence using the SFR and stellar mass estimates from \pros. 

\subsection{Determining stellar mass dependence using Prospector SED fits}  \label{section:massdependence_firstsubsec}
For this analysis, we focus on the $z < 1$ SWIRE-selected galaxy sample with acceptable \pros\ SED fits. We excluded sources that showed a significant AGN contribution (either radiative mode or radio excess, as defined in Section\,\ref{section:final_classifications}). We also removed the radio non-detected sources that were flagged as AGN using spectroscopy, X-ray detections, or mid-IR colour-colour cuts by prior studies \citep{Duncan2021}. After applying these cuts, we were left with a sample of 85,162 predominantly star-forming and passive galaxies. Following \citetalias{Smith2021}, we extend the stacked PDF method described in Section\,\ref{section:radio_excess} to three dimensions. 100 samples are generated for each source along the SFR, stellar mass, and $L_\mathrm{150~MHz}$ axes. We adopt skewed error distributions in log space for SFR and stellar mass using the 16\tsu{th}, 50\tsu{th}, and 84\tsu{th} percentile \pros\,estimates. $L_\mathrm{150~MHz}$ is sampled from a normal distribution in linear space centred at the maximum likelihood estimate and with standard deviation equal to the RMS. 

We then create a three-dimensional histogram of the samples, summing over all the sources to create a stacked PDF. We use 50 equally spaced logarithmic stellar mass bins between $7.5 < \log_{10}(M/M_\odot) < 11.8$, 60 equally spaced logarithmic SFR bins between $-3 < \log_{10}(\psi/M_\odot~\mathrm{yr}^{-1}) < 3$, and 180 equally spaced logarithmic $L_\mathrm{150~MHz}$ bins between $17 < \log_{10}(L_\mathrm{150~MHz}/\mathrm{W~Hz}^{-1}) < 26$ and calculate the median $L_\mathrm{150~MHz}$ in each SFR and stellar mass bin. As before, the samples with $L_\mathrm{150~MHz} < 10^{17}~\mathrm{W~Hz}^{-1}$ and SFR $< 10^{-3}~M_\odot~\mathrm{yr}^{-1}$ are arbitrarily assigned to the lowest $L_\mathrm{150~MHz}$ and log SFR bins, respectively. 

\begin{figure*}
\centering
\includegraphics[width=1\linewidth]{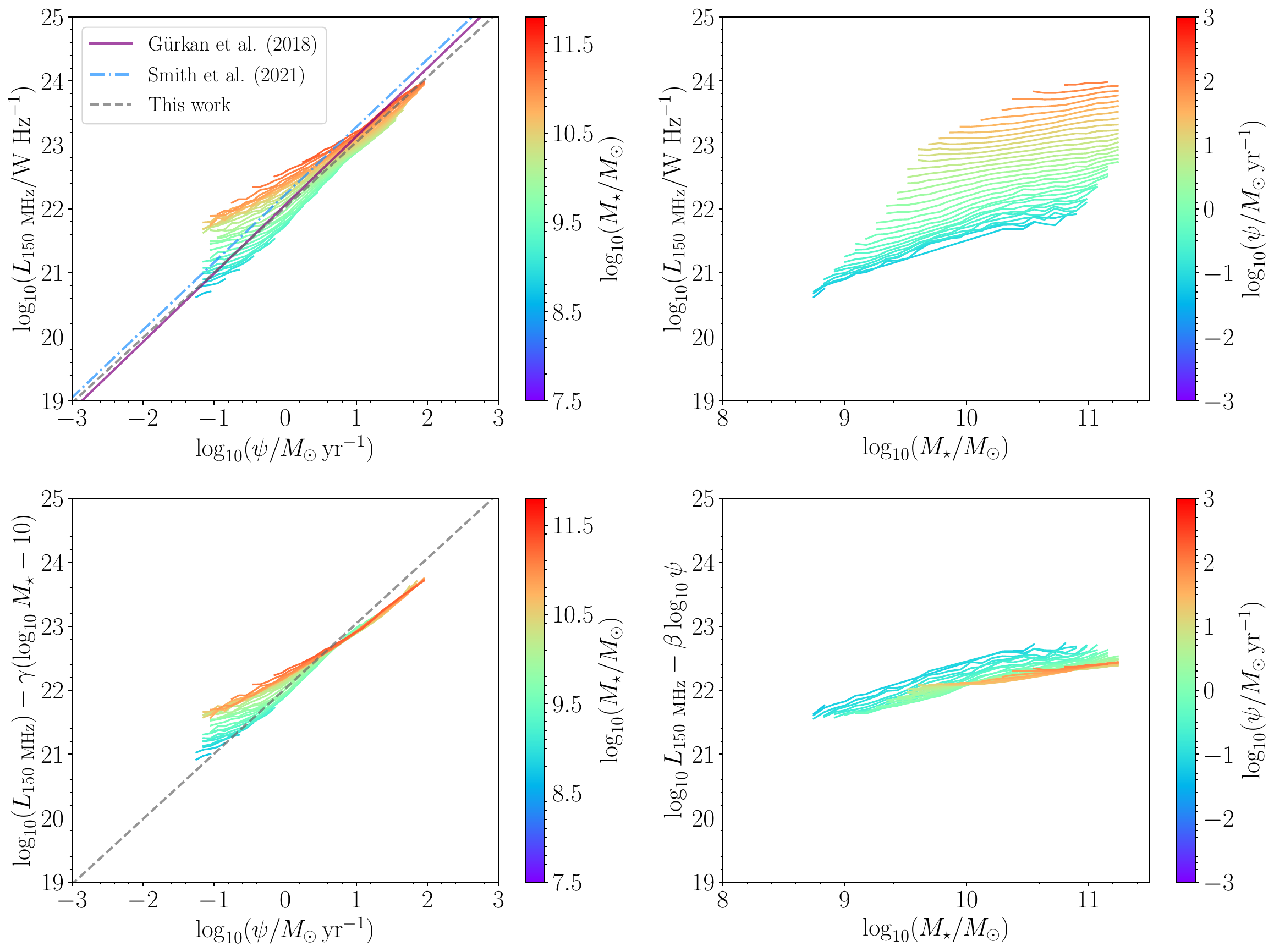}
\caption{The relationship between radio luminosity at 150 MHz and the dependence on stellar mass and star formation rate (SFR), derived for the 85,162 $z<1$ SWIRE-selected sources excluding known/likely AGN. Stellar mass and SFRs are estimated through \pros\,SED fits. In the top left-hand panel, the median likelihood 150 MHz radio luminosity - SFR plot is shown for different stellar mass bins. The different coloured lines represent distinct stellar mass bins, as indicated in the colour-bar to the right. The top right-hand panel illustrates the median likelihood stellar mass vs 150 MHz radio luminosity plots for different SFR bins, with colours again relative to the bar to the right. Similarly, in the bottom left-hand panel, the stellar mass-adjusted radio luminosity - SFR plot is displayed for varying stellar mass bins, using the best fit values from Equation\,\ref{eq:massdep}. Finally, the bottom right-hand panel exhibits the SFR-adjusted radio luminosity vs stellar mass plot for different SFR bins. }
\label{figure:mass_dependence_all}
\end{figure*}

Similar to \citetalias{Gurkan2018} and \citetalias{Smith2021}, we find clear evidence of stellar mass dependence in the SFR--$L_\mathrm{150~MHz}$ relation (Figure\,\ref{figure:mass_dependence_all}). In all the panels of Figure\,\ref{figure:mass_dependence_all}, only the bins populated by a minimum of 15 galaxies are displayed, taking into account that each source is sampled 100 times. The top-left panel of this figure shows the median-likelihood $L_\mathrm{150~MHz}$ as a function of SFR in different stellar mass bins indicated by the colour bar to the right. Interestingly, this plot indicates that the relationship between SFR and $L_\mathrm{150~MHz}$ becomes progressively sub-linear as we move towards higher stellar mass bins. This observation may suggest the influence of non-calorimetric processes, such as the rapid diffusion of CREs within star-forming regions, as discussed by \citet{Berkhuijsen2013} and \citet{Heesen2019}, or perhaps the presence of undiagnosed low-luminosity contamination by AGN \citepalias[as discussed in][]{Smith2021}. Furthermore, in contrast to \citetalias{Smith2021} our \pros\ results indicate that the stellar mass dependence (as apparent by the difference between the most- and least-150\,MHz luminous stellar mass bins), appears weaker at $\log_{10} (\psi/M_\odot~\mathrm{yr}^{-1}) > 0.5$ than for the lower SFR galaxies. 

The top-right panel of Figure\,\ref{figure:mass_dependence_all} displays the median-likelihood $L_\mathrm{150~MHz}$ as a function of stellar mass in different SFR bins. It is immediately clear that there is an increase in $L_\mathrm{150~MHz}$ with stellar mass (consistent with \citetalias{Smith2021} and \citealt{Heesen2022}). However, in keeping with expectations on the basis of the top-left panel, the gradient of the stellar mass--radio luminosity relation appears to flatten with increasing stellar mass (though the increasingly small stellar mass range probed for the highest SFRs precludes a definitive statement based on visual inspection alone). 

To investigate these trends further, we fit our data using the stellar mass dependent form of the SFR--$L_\mathrm{150~MHz}$ relation from \citetalias{Gurkan2018}:

\begin{equation} 
\label{eq:massdep}
  L_\mathrm{150~MHz} = L_1~\psi^\beta~\left(\frac{M}{10^{10}M_\odot}\right)^\gamma.
\end{equation}

\noindent We use \emcee\,with 16 walkers and 10,000 iterations to calculate the uncertainties associated with these fits. The following best fit values are obtained: $\log_{10}L_1 = 22.083\pm0.004$, $\beta = 0.778\pm0.004$, and $\gamma = 0.334\pm0.006$. \citetalias{Smith2021} used simulations to reveal small biases in these estimates, and suggested that the uncertainties are likely to be underestimated (with true uncertainties around $\sigma_\beta = 0.011, \sigma_{\log_{10}L_1} = 0.016$, and $\sigma_\gamma = 0.037$; correcting for these biases will be discussed at length in Shenoy et al. \textit{in preparation}). By way of comparison, \citetalias{Smith2021} obtained best fit estimates of $\log_{10}L_1 = 22.218\pm0.016$, $\beta = 0.903\pm0.012$, and $\gamma = 0.332\pm0.037$, while \citetalias{Gurkan2018} found $\log_{10}L_1 = 22.13\pm0.01$, $\beta = 0.77\pm0.01$, and $\gamma = 0.43\pm0.01$. While the exponent of SFR dependence ($\beta$) seen in our results is close to that found by \citetalias{Gurkan2018},  the exponent of the stellar mass dependence differs. On the other hand, we see a similar degree of stellar mass dependence to \citetalias{Smith2021}, but the SFR dependence differs between the two works. To determine whether incompleteness in our sample selection could be playing a role in our findings, we followed \citetalias{Smith2021} and repeated our analyses using a 95 per cent stellar mass-complete sample of 64,838 sources derived using the method of \citet{Pozzetti2010}, finding that our results are statistically unchanged. We therefore conclude that stellar mass completeness does not significantly impact our results.

In the bottom-left panel of Figure\,\ref{figure:mass_dependence_all}, we plot the stellar mass-adjusted radio luminosity (obtained by dividing the $L_\mathrm{150~MHz}$ data shown in the top panels by the best fit stellar mass-dependent term in Equation\,\ref{eq:massdep}) in an attempt to reveal the SFR--$L_\mathrm{150~MHz}$ relationship with the stellar mass effects accounted for. At $\log_{10} (\psi/M_\odot~\mathrm{yr}^{-1}) > 0.5$, the $L_\mathrm{150\,MHz}$ values in different stellar mass bins virtually collapse into a single line, while at lower SFRs there is still significant scatter, to a similar degree seen in \citetalias{Smith2021}. The differences observed between our results and those of \citetalias{Smith2021}, as well as the inferred stellar mass dependence in the SFR--$L_\mathrm{150~MHz}$ relationship, may be partially attributed to the assumed SFH and the various explicit or implicit priors used in the SED fitting process. While the non-parametric continuity prior used in this work has been shown to be flexible enough for modelling a diverse range of galaxy SEDs \citep[Das et al. \textit{in preparation}]{Leja2019B,Johnson2021,Haskell2023B}, it is still by no means perfect and may not be uniformly appropriate at all stellar mass scales \citep[e.g. lower mass galaxies are expected to have burstier SFHs;][]{Hopkins2023}. Moreover, the apparent `break' in the SFR--$L_\mathrm{150~MHz}$ relation might suggest the requirement for additional physics.
  
\subsection{The influence of passive galaxies}

Our findings in Section\,\ref{section:massdependence_firstsubsec} point towards a more pronounced dependence on stellar mass in the SFR--$L_\mathrm{150~MHz}$ relationship at $\log_{10}\mathrm{(SFR)} < {0.5} M_\odot\mathrm{yr}^{-1}$. To investigate a possible cause, we repeat our previous analysis after excluding any quiescent galaxies, which we identify as follows. For each source, we calculate the SFR expected based on the parametrisation of the star-formation rate -- stellar mass relation from \citet{Schreiber2015}, and identify those 7,081 galaxies with \pros\ SFR lower than $<0.6\,$dex below the relation as passive galaxies\footnote{We choose a looser cutoff than the typical $\pm 0.3$ dex scatter associated with the MS \citep[e.g.][]{Tacchella2022A,Haskell2023A} in order to account for the large uncertainties associated with very low SFRs.}. The results appear similar -- and the best-fit parameters obtained using Equation\,\ref{eq:massdep} to fit the data after correcting for biases ($\log_{10}L_1 = 22.085\pm0.004$, $\beta = 0.783\pm0.005$, and $\gamma = 0.338\pm0.006$) are consistent with the values derived in Section\,\ref{section:massdependence_firstsubsec}. This is not unexpected since the star-forming galaxy population is vastly numerically dominant, and we are using median-likelihood $L_\mathrm{150\,MHz}$ values in each bin. Excluding the 7,081 quiescent galaxies in this way results in an apparently reduced scatter in the 150\,MHz radio luminosity--SFR relation. The apparent reduction in scatter, however, can be attributed purely to the smaller sample size (since fewer bins meet the minimum occupancy threshold of 15 galaxies, especially at low SFRs) rather than a difference in the relationship between luminosity and SFR for passive galaxies, for example.
 
\section{Conclusions} \label{section:conclusions}
In this work, we have used \pros\ \citep{Leja2017,Johnson2021} to estimate the physical properties of sources over 6\,deg$^2$ of the ELAIS-N1 field. \pros\,offers substantial advantages compared to other SED fitting codes, namely, the inclusion of AGN on an even footing with the stellar population synthesis, the implementation of non-parametric star formation histories, the modelling of nebular emission lines and the use of a dynamic nested sampling framework to efficiently sample multi-modal parameter spaces and produce realistic uncertainties alongside the source properties. The main conclusions of this work are as follows:
\begin{enumerate}
    \item We successfully produce radio source classifications for 92 per cent of the 31,610 150\,MHz sources in the ELAIS-N1 field of the LoTSS first data release. Each source is classified as either a star-forming galaxy, radio-quiet AGN, high-excitation radio galaxy (HERG) or low-excitation radio galaxy (LERG).
    \item We have shown that our classifications are of a quality which is similar to those produced using four codes in a previous work. This was possible because the inclusion of AGN models in \pros\,-- on an even footing with the stellar population synthesis -- enables us to build on previous work \citep[e.g.][]{Best2023} by identifying sources that are the likely hosts of radiative mode AGN, as well as those for which there is a radio excess above that expected on the basis of the \pros\,SFR. 
    \item We used \pros\,to estimate the physical properties of 133,000 3.6\,$\mu$m-selected sources, and revisited the relationship between a source's SFR and its radio luminosity. 
    \item Our best-fit SFR--$L_\mathrm{150~MHz}$ relations are:
    \begin{itemize}
        \item $\log_{10}L_\mathrm{150~MHz} = (22.024\pm0.006) + (1.019\pm0.009)~\log_{10}(\psi/M_\odot\,\mathrm{yr}^{-1})$ for the stellar mass independent case, and \item $\log_{10}~L_\mathrm{150~MHz} = (22.083\pm0.004) + (0.778\pm0.004) \log_{10}(\psi/M_\odot\,\mathrm{yr}^{-1}) + (0.334\pm0.006) \log_{10} (M/10^{10}M_\odot)$ if we assume the stellar mass-dependent form adopted in previous works (e.g. \citetalias{Gurkan2018} and \citetalias{Smith2021}). 
    \end{itemize}
    \item We found that, using \pros\ to fit exactly the same input photometry as used by previous works (e.g. \citetalias{Smith2021}), the form of the stellar mass dependence in the SFR--$L_\mathrm{150\,MHz}$ relation differs from what had previously been observed using \magp\ for the SED fitting. The \pros\ results indicate a far-tighter relation, with lower scatter especially at $\log_{10} (\psi/M_\odot\,\mathrm{yr}^{-1}) > 0.5$ when the stellar mass effects are accounted for. 
\end{enumerate}

This work has focused on SED modelling using only photometric data, and it is clear that even using all available codes and every photometric band there can still be significant doubt over what the ``correct'' classification is for any particular radio source. The most reliable radio source classifications require spectroscopy \citep[e.g.][Drake et al. \textit{in preparation}]{Best2012}. In this context, it is timely that the WEAVE instrument \citep{Dalton2016,Jin2023} has started collecting data, and the initial batch of over a million spectra of LOFAR-selected sources will be available in the coming months as part of the WEAVE-LOFAR survey \citep{Smith2016}. This will ultimately include every source in the ELAIS-N1 field brighter than $\sim$100 $\mu$Jy. \pros\,is uniquely placed to deal with the WEAVE-LOFAR data since it can deal with photometry and spectra on an equal footing. We are hopeful that including the WEAVE spectra in our modelling will allow us to put stronger constraints on the physical parameters (e.g. SFRs) obtained for ELAIS-N1 sources, and significantly improve the accuracy of galaxy classifications. There are great opportunities for this kind of galaxy analysis given the range of new and forthcoming data sets in the era of massively-multiplexed spectroscopic surveys such as 4MOST \citep{deJong2019}, Subaru-PFS \citep{Greene2022} and MOONS \citep{Cirasuolo2016}, for studying new samples from e.g. \textit{Euclid}, LMT\slash TolTEC \citep{Pope2019,Wilson2020} as well as the SKA \citep{Rawlings2004}.

\section*{Acknowledgements}
We thank the anonymous referee for their insightful comments. SD acknowledges support from a Science and Technology Facilities Council (STFC) studentship via grant ST/W507490/1. DJBS, MJH and ABD acknowledge support from UK STFC under grant ST/V000624/1. KJD acknowledges funding from the European Union’s Horizon 2020 research and innovation programme under the Marie Sk{\l}odowska-Curie grant agreement No. 892117 (HIZRAD) and support from the STFC through an Ernest Rutherford Fellowship (grant number ST/W003120/1). PNB and RK are grateful for support from the UK STFC via grant ST/V000594/1. MIA and SS acknowledge support from the UK STFC (grant numbers ST/V506709/1 and ST/X508408/1, respectively). KM has been supported by the National Science Centre (UMO-2018/30/E/ST9/00082). LKM is grateful for support from the Medical Research Council [MR/T042842/1]. IP acknowledges support from INAF under the Large Grant 2022 funding scheme (project ``MeerKAT and LOFAR Team up: a Unique Radio Window on Galaxy/AGN co-Evolution''). LOFAR is the Low Frequency Array designed and constructed by ASTRON. It has observing, data processing, and data storage facilities in several countries, which are owned by various parties (each with their own funding sources), and that are collectively operated by the ILT foundation under a joint scientific policy. The ILT resources have benefited from the following recent major funding sources: CNRS-INSU, Observatoire de Paris and Université d’Orléans, France; BMBF, MIWFNRW, MPG, Germany; Science Foundation Ireland (SFI), Department of Business, Enterprise and Innovation (DBEI), Ireland; NWO, The Netherlands; The Science and Technology Facilities Council, UK; Ministry of Science and Higher Education, Poland; The Istituto Nazionale di Astrofisica (INAF), Italy. This research made use of the Dutch national e-infrastructure with support of the SURF Cooperative (e-infra 180169) and the LOFAR e-infra group. The Jülich LOFAR Long Term Archive and the German LOFAR network are both coordinated and operated by the Jülich Supercomputing Centre (JSC), and computing resources on the supercomputer JUWELS at JSC were provided by the Gauss Centre for Supercomputing e.V. (grant CHTB00) through the John von Neumann Institute for Computing (NIC). This research made use of the University of Hertfordshire high performance computing facility and the LOFAR-UK computing facility located at the University of Hertfordshire and supported by STFC (ST/V002414/1), and of the Italian LOFAR IT computing infrastructure supported and operated by INAF, and by the Physics Department of Turin University (under an agreement with Consorzio Interuniversitario per la Fisica Spaziale) at the C3S Supercomputing Centre, Italy. For the purpose of open access, the author has applied a Creative Commons Attribution (CC BY) licence to any Author Accepted Manuscript version arising from this submission.

%%%%%%%%%%%%%%%%%%%%%%%%%%%%%%%%%%%%%%%%%%%%%%%%%%
\section*{Data Availability}

The multiwavelength catalogues used in this work are available at the LOFAR surveys website at \url{https://lofar-surveys.org/deepfields.html}. Output catalogues with parameter estimates (including stellar mass, SFR, dust luminosity, and AGN fraction) and a table of classifications is made available at \url{https://lofar-surveys.org/deepfields.html} as part of this paper. More extensive SED-fitting results pertaining to this paper will be shared on reasonable request to the first author.

%%%%%%%%%%%%%%%%%%%% REFERENCES %%%%%%%%%%%%%%%%%%

% The best way to enter references is to use BibTeX:

\bibliographystyle{mnras}
\bibliography{bibfile} % if your bibtex file is called example.bib

% Alternatively you could enter them by hand, like this:
% This method is tedious and prone to error if you have lots of references
%\begin{thebibliography}{99}
%\bibitem[\protect\citeauthoryear{Author}{2012}]{Author2012}
%Author A.~N., 2013, Journal of Improbable Astronomy, 1, 1
%\bibitem[\protect\citeauthoryear{Others}{2013}]{Others2013}
%Others S., 2012, Journal of Interesting Stuff, 17, 198
%\end{thebibliography}

%%%%%%%%%%%%%%%%%%%%%%%%%%%%%%%%%%%%%%%%%%%%%%%%%%

%%%%%%%%%%%%%%%%% APPENDICES %%%%%%%%%%%%%%%%%%%%%

\appendix

\section{Assessment of Prospector derived SFRs}

\subsection{Comparison with spectroscopically measured SFRs} \label{appendix:sdsssfr}

\begin{figure}
\centering
\includegraphics[width=0.8\linewidth]{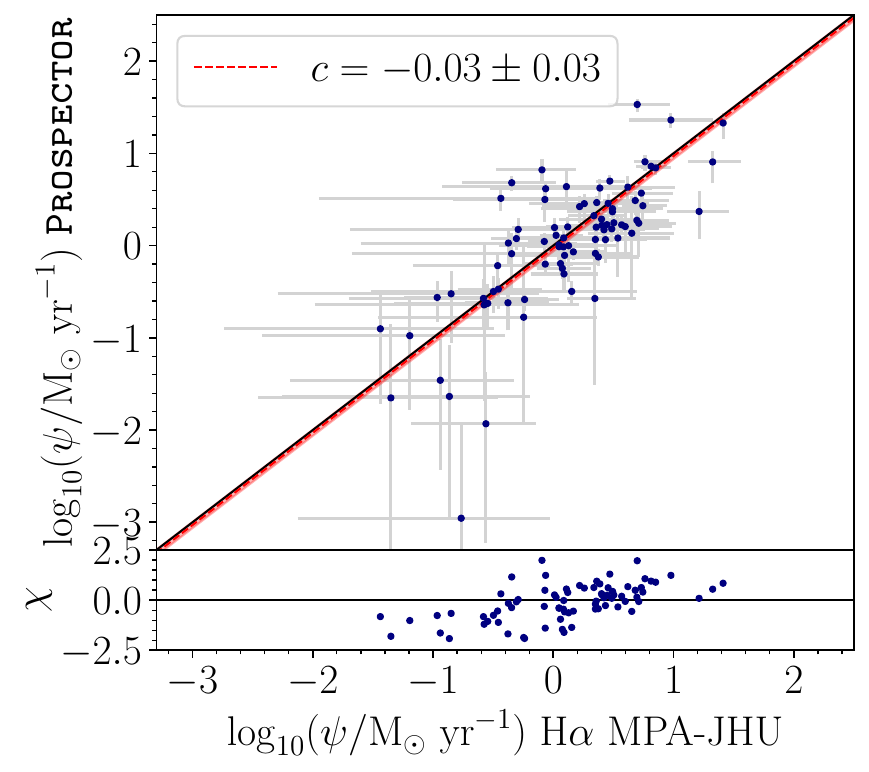}
\caption{Comparison of \pros\ estimated median likelihood SFRs (averaged over 10 Myrs) with the Balmer corrected H$\alpha$ SFRs. The solid black line denotes the ideal 1:1 relation. The red dashed line represents the best-fit deviation of the \pros\ estimated SFRs from the H$\alpha$ SFRs, while the bottom panel shows the difference of the \pros\ SFR estimates of the individual sources from H$\alpha$ SFRs, in units of the (propagated) uncertainties.}
\label{fig:mpajhu-vs-conti-sfr}
\end{figure}

To determine the reliability of SFRs obtained using SED fitting, we compare \pros\ \citep{Johnson2021} SFR estimates with those obtained from H$\alpha$ emission line measurements. The latter, which accounts for the possible presence of dust, is widely regarded as the gold standard tracer of SFR up to 10 Myrs \citep[see][]{Calzetti2013,Tacchella2022C}. We use the SFRs measured from the dust extinction and aperture corrected Sloan Digital Sky Survey \citep[SDSS DR8;][]{Abazajian2009} line fluxes provided by the Max Planck Institute for Astrophysics and the Johns Hopkins University (MPA-JHU) group \citep[see][]{Brinchmann2004}. We match the optical positions of sources in our SWIRE-selected sample with the MPA-JHU catalogue using a nearest-neighbour algorithm with a maximum 1 arcsec separation, finding 241 common sources. Our analysis is limited to 96 sources not flagged as stars by the Schlegel classification \citep{Bolton2012}, with significant ($> 3\sigma$) H$\beta$ detection (allowing reliable dust corrections) and acceptable \pros\ fits. For this test, we average \pros\,SFRs over 10\,Myrs to enable a like-for-like comparison given the expected timescale associated with the Balmer line emission \citep[e.g.][]{Kennicutt2012}. Figure \ref{fig:mpajhu-vs-conti-sfr} shows excellent consistency between the \pros\ and Balmer-line SFRs. We obtain a best-fit relation of $\log_{10}(\psi_\mathrm{\pros}) = \log_{10}(\psi_\mathrm{H\alpha}) - (0.03\pm0.03)$, as well as a reduced $\chi^2 \approx 1$ (where $\chi = \frac{\psi_\mathrm{\pros} - \psi_\mathrm{H\alpha}}{\sigma_\psi}$; $\sigma_\psi$ represents the \pros\ and H$\alpha$ SFR uncertainties added in quadrature).

\subsection{Star forming main sequence} \label{appendix:mainseq}
\begin{figure}
\centering
\includegraphics[width=0.95\linewidth]{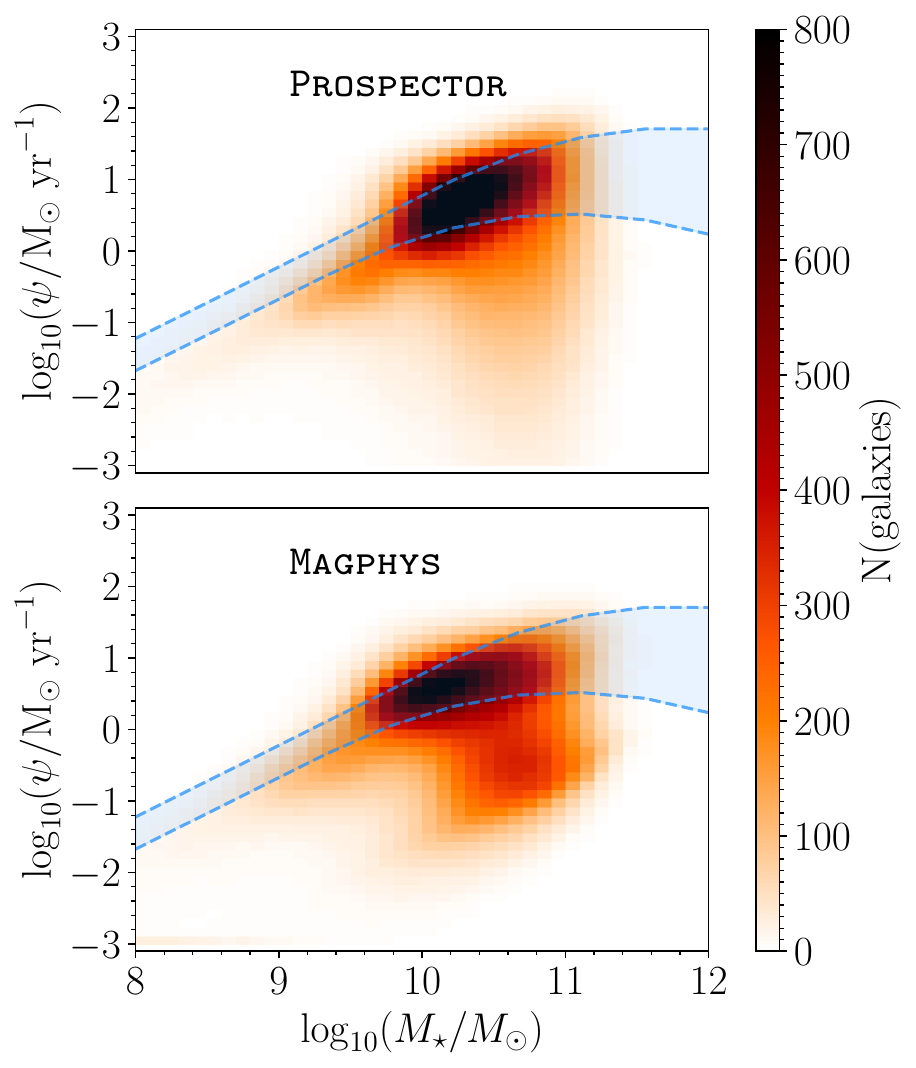}
\caption{Stellar mass vs SFR plots for $\sim$120,000 SWIRE-selected sources using estimates obtained from \pros\ (top panel) and \magp\ (bottom panel) SED fits. The blue shaded region represents the region enclosed by the \citet{Schreiber2015} parametrisation of the main sequence of star forming galaxies between $0 < z < 1$, converted to our adopted \citet{Kroupa2001} IMF.}
\label{fig:mainseq}
\end{figure}

To further assess the performance of \pros, in Figure \ref{fig:mainseq} we show a comparison  between the location of our SWIRE-selected sample in the stellar mass--SFR plane, derived using the  \pros\ (top) and \magp\ (bottom) SED fits. We focus on the 120,307 sources from the complete SWIRE-selected sample for which \pros\ produced acceptable fits. To produce the PDFs, we follow an approach similar to the one used in Section \ref{section:radio_excess}, wherein we generate 100 samples from the stellar mass and SFR distributions for each source using 50 bins evenly spaced between $7 < \log_{10} (M_\star \slash M_\odot) < 12$, and $-3 < \log_{10} (\psi \slash  M_\odot\,\mathrm{yr}^{-1}) < 3$, respectively. Sources with $\log_{10}(\psi/M_\odot~\mathrm{yr}^{-1}) < -3$ were arbitrarily assigned to the lowest log SFR bins. The PDFs shown are derived by summing the individual stellar mass--SFR PDFs over the whole sample. 

In both panels we have overlaid a shaded region enclosing the location of the  ``main sequence'' (hereafter MS) expected for star forming galaxies at $0 < z < 1$ using the \citet{Schreiber2015} parameterisation, and converted to our adopted IMF. In each case, a distinct MS is evident, accompanied by a region of quiescent galaxies (i.e. those falling below the main sequence), in line with our expectations on the basis of recent literature \citep[e.g.][]{Brinchmann2004, Daddi2007, Elbaz2007, Whitaker2012}. Together with the results in Appendix \ref{appendix:sdsssfr}, these results offer significant encouragement that our \pros\ SED fitting is producing realistic results. 

%%%%%%%%%%%%%%%%%%%%%%%%%%%%%%%%%%%%%%%%%%%%%%%%%%

% Don't change these lines
\bsp  % typesetting comment
\label{lastpage}
\end{document}